\documentclass[12pt,a4paper]{article} 
\pdfoutput=1
\usepackage{graphicx}
\usepackage{jheppub}

\usepackage{amsmath,amssymb,bm,amsfonts,yfonts}

\newcommand{\be}{\begin{equation}}
\newcommand{\ee}{\end{equation}}
\newcommand{\beqa}{\begin{eqnarray}}
\newcommand{\eeqa}{\end{eqnarray}}
 \newcommand{\bn}{\begin{enumerate}}
\newcommand{\en}{\end{enumerate}}

\def\bra#1{\left\langle #1\right|}
\def\eeq{\end{equation}}

\def\ket#1{\left| #1\right\rangle}

\def\Tr{\mathop{\rm Tr}}

\relax

\title{Fluctuations around Bjorken Flow and the onset of turbulent phenomena}

\author{Stefan Floerchinger and Urs Achim Wiedemann
\\
\vspace{0.1in}

Physics Department, Theory Unit, CERN, CH-1211 Gen\`eve 23, Switzerland
\vspace{0.1in}

E-mail addresses: {\tt Stefan.Floerchinger@cern.ch, Urs.Wiedemann@cern.ch}
}

\abstract{
We study how fluctuations in fluid dynamic fields can be dissipated or amplified within the characteristic spatio-temporal structure of a heavy ion collision. The initial conditions for a fluid dynamic evolution of heavy ion collisions may contain significant fluctuations in all fluid dynamical fields, including the velocity field and its vorticity components.  We formulate and analyze the theory of local fluctuations around average fluid fields described by Bjorken's model. For conditions of laminar flow, when a linearized treatment of the dynamic evolution applies, we discuss explicitly how fluctuations of large wave number get dissipated while modes of sufficiently long wave-length pass almost unattenuated or can even be amplified.
In the opposite case of large Reynold's numbers (which is inverse to viscosity), we establish that (after suitable coordinate transformations) the dynamics is governed by an evolution equation of non-relativistic Navier-Stokes type that becomes essentially two-dimensional at late times. One can then use the theory of Kolmogorov and Kraichnan for an explicit characterization of turbulent phenomena in terms of the  wave-mode dependence of correlations of fluid dynamic fields. We note in particular that fluid dynamic correlations introduce characteristic power-law dependences in two-particle correlation functions. 
}

\begin{document}
\def\vev#1{\langle#1\rangle}
\def\ov{\over}
\def\le{\left}
\def\ri{\right}
\def\ha{{1\over 2}}
\def\lam{{\lambda}}
\def\Lam{{\Lambda}}
\def\al{{\alpha}}
\def\ket#1{|#1\rangle}
\def\bra#1{\langle#1|}
\def\vev#1{\langle#1\rangle}
\def\det{{\rm det}}
\def\tr{{\rm tr}}
\def\Tr{{\rm Tr}}
\def\NN{{\cal N}}
\def\Om{{\Omega}}
\def\th{{\theta}}
\def\lam {\lambda}
\def\om {\omega}
\def\ra {\rightarrow}
\def\ga {\gamma}
\def\sig{{\sigma}}
\def\ep{{\epsilon}}
\def\apr{{\alpha'}}
\newcommand{\p}{\partial}
\def\LL{{\cal L}}
\def\HH{{\cal H}}
\def\GG{{\cal G}}
\def\TT{{\cal T}}
\def\CC{{\cal C}}
\def\OO{{\cal O}}
\def\PP{{\cal P}}
\def\tir{{\tilde r}}

\newcommand{\bea}{\begin{eqnarray}}
\newcommand{\eea}{\end{eqnarray}}
\newcommand{\nn}{\nonumber\\}

\maketitle

\section{Introduction}
\label{sec1}
It is a long-standing idea, first articulated by Landau in the 1950s, that the evolution of
matter compressed in nuclear collisions lends itself to a fluid dynamical description~\cite{Landau:1953gs}. 
On the experimental side, characteristic correlations of particle production with the event 
plane had been interpreted as qualitative support for fluid dynamic behavior since the very first 
relativistic heavy ion collision experiments at the BEVALAC in the 1980s \cite{Gustafsson:1984ka}. 
Early qualitative predictions, based on fluid dynamics, 
include notably the argument~\cite{Ollitrault:1992bk} that the second harmonics of the azimuthal 
particle distribution (elliptic flow $v_2$) changes at mid-rapidity from out-of-plane to in-plane emission 
at higher center of mass energy. This was confirmed experimentally 
in heavy ion collisions at the Brookhaven National Laboratory (BNL) Alternating Gradient Synchrotron
($\sqrt{s_{\rm NN}} < 5$ GeV), and at the CERN Super Proton Synchrotron
($\sqrt{s_{\rm NN}} < 20$ GeV), for a review see e.g. Refs.~\cite{Ollitrault:1997vz}.
Also, fluid dynamic arguments provided early on some 
qualitative understanding of the dependence of elliptic flow on particle species, and on the 
energy and rapidity dependence of the collective sidewards displacement of particle production 
at projectile rapidity (sidewards flow $v_1$)~\cite{Ollitrault:1997vz}.
However, conclusions remained largely limited to the qualitative statement that the observed flow 
in semi-central collisions ``retains some signature of the pressure in the high density region created 
during the initial collision''~\cite{Appelshauser:1997dg}. 

This changed soon after  the start of the  Relativistic Heavy Ion Collider (RHIC) in the year 2000,
when several  groups~\cite{Teaney:2000cw,Teaney:2001av,Kolb:2000fha,Kolb:2001qz} noted  
that fluid dynamic simulations of Au+Au collisions at $\sqrt{s_{\rm NN}} < 200$ GeV 
can account  {\it quantitatively} for the main manifestations of collectivity at RHIC, 
including the dominant elliptic flow signal at mid rapidity and its dependencies on transverse 
momentum, centrality and particle species. These studies were based on 
simplified  2+1-dimensional simulations, following Bjorken's argument
that the initial conditions for fluid dynamic fields are close to longitudinally boost-invariant, 
and that this boost-invariance is preserved by the fluid dynamics~\cite{Bjorken:1982qr}.
Moreover, early comparisons to RHIC data relied on ideal fluid dynamic equations of motion 
without dissipative effects. Soon afterwards, Teaney~\cite{Teaney:2003kp} observed in 2003 
that even very small values of the ratio $\eta/s$ of shear viscosity over entropy density induce 
dissipative effects that result in a sizable reduction of the elliptic flow signal. Therefore, to the
extent to which uncertainties in the comparison of fluid dynamic simulations with data can be
controlled quantitatively, measurements of collective flow in heavy ion collisions provide a 
sensitive tool for constraining transport properties of QCD matter. This is one of the main
motivations for the development of more and more detailed fluid dynamic simulations of 
relativistic heavy ion collisions in recent years, see 
e.g.~\cite{Heinz:2009xj,Romatschke:2009im,Teaney:2009qa,Hirano:2007gc} for recent reviews.

Ideal fluid dynamics
is determined fully by the equation of state and conservation laws. For causal viscous fluid dynamics,  
transport properties and relaxation times enter in addition. But the equation of state, 
transport properties and relaxation times are in principle calculable from first principles of a given quantum field theory. Therefore, a comparison of fluid dynamic 
simulations to data of heavy ion collisions has the potential of constraining properties of QCD 
matter that are fundamental in the sense that they are most directly related to the QCD lagrangian. 
In practice, the bottleneck for such a program is the limited control over the initial data that
are evolved fluid dynamically~\footnote{In principle, both the initial conditions for fluid dynamic evolution as well as the conditions for decoupling from the fluid dynamic evolution ('freeze-out') imply assumptions. However, fluid dynamic evolution occurs in response to pressure gradients that are much larger at initial times. This indicates that the modeling of freeze-out and final decoupling does not presently limit the predictive power of the approach.}.
Early comparisons of fluid dynamic simulations with RHIC data~\cite{Teaney:2000cw,Teaney:2001av,Kolb:2000fha,Kolb:2001qz} employed
a set of smooth, event-averaged initial conditions that were specified via the average
collision geometry in terms of a transverse energy (or entropy) distribution with vanishing 
flow at initial times.  Even within this limited set of initial conditions, one observed that differences 
in the initial transverse profile of phenomenologically motivated models could result in variations 
of the dominant elliptic flow signal by up to 30\%~\cite{Hirano:2005xf,Romatschke:2007mq}.

Within recent years, there has been a growing realization of the importance of event-by-event
fluctuations in constraining the initial conditions of fluid dynamic evolution in heavy ion collisions. 
In particular, event-averaged initial conditions reflect the symmetries of the almond-shaped nuclear
overlap region of finite impact parameter collisions and can therefore give rise only to dipole, 
quadrupole and higher even moments in the initial density distribution. In marked contrast, the azimuthal 
momentum distributions measured by all experiments at the 
LHC~\cite{ALICEflow,ALICEfluct,ATLASflow,CMSflow} and at RHIC~\cite{STARflow,PHENIXflow}
show a prominent third harmonic moment $v_3$, as well as non-vanishing moments  
$v_1$ and $v_5$ in addition to the expected even ones. These structures had been attributed
previously to other speculative effects ("Mach-cone", "ridge"), but as pointed out first by Alver and 
Roland~\cite{Alver:2010gr} (see also Sorensen~\cite{Sorensen:2010zq} for a related earlier suggestion)
they emerge most naturally from the fluid dynamical evolution of initial density inhomogeneities.
In addition, fluctuations increase the spatial eccentricity of initial transverse density distributions, and
this accounts naturally for the fact that elliptic flow values remain sizable in the most central
collisions and for smaller colliding systems~\cite{Alver:2008zza}. There is by now compelling 
evidence that the dynamical evolution of fluctuating initial conditions is a prerequisite
for a detailed quantitative understanding of flow in heavy ion 
collisions~\cite{Luzum:2011mm,Schenke:2011qd}. And since the various
flow moments $v_n$ depend differently on the event-averaged initial state and its
event-by-event fluctuations, analyzing the dynamical evolution of these initial fluctuations provides
a novel tool for constraining the main uncertainty in fluid dynamical simulations of heavy ion collisions.

There has been a significant effort recently in studying fluctuating initial conditions in heavy ion collisions
~\cite{Broniowski:2007ft,Hirano:2009bd,Staig:2010pn,Staig:2011wj,Mocsy:2010um,Heinzmore}
and studying their propagation in full fluid dynamic simulations or transport models
\cite{Takahashi:2009na,Werner:2010aa,Petersen:2010md,Petersen:2010cw,Qin:2010pf,Holopainen:2010gz,Qiu:2011iv,Schenke:2010rr,Alver:2010dn}.
Precursors of these developments include e.g. Refs.~\cite{Gyulassy:1996br,Miller:2003kd,Socolowski:2004hw}.
The recent efforts focussed mainly on initial density inhomogeneities. 
But more general fluctuating initial conditions are conceivable. For instance, (non fluid-dynamical) initial 
fluctuations in the flow field $u^\mu$ may be expected to accompany a fluctuating initial energy density 
profile $\epsilon$~\cite{Petersen:2010cw}.
Even if fluctuations in the initial spatial distribution may turn out to be sufficient to account for the 
measured flow components, it is clearly important to constrain such other conceivable sources of 
initial fluctuations since these may confound any quantitative interpretation of flow phenomena
aimed at an extraction of $\eta/s$ and other fundamental properties of QCD matter. 
This argues for treating fluctuations in all fluid dynamical
fields democratically. 

The present paper aims at supplementing the current discussion of fluctuating initial
conditions with a model study in which the propagation of initial fluctuations can be 
followed in a very explicit, partly analytical way. To this end, we formulate the fluid dynamic 
evolution of fluctuations in {\it all} fluid dynamic fields around an event-averaged Bjorken flow profile. 
The inclusion of fluctuations in all fields will provide access  to qualitatively novel features
such as the dynamical evolution of vorticity. It will also allow us to discuss anew
how one of the most characteristic manifestations of fluid dynamics, namely turbulence, can emerge in the 
specific expanding geometry of a Bjorken-like flow profile. The issue here is not whether heavy ion 
collisions can display fully developed turbulence: It has been pointed out previously
(see e.g. Ref.~\cite{Romatschke:2007eb}) 
that the relevant Reynolds numbers are typically larger than unity, since they are proportional to the inverse of
the normalized viscosity $\eta/s$. However, the length and time-scales in heavy ion collisions
are so small that ${\rm Re} < {\cal O}(100)$ which is well below the conditions under which fully developed 
turbulence is expected. Rather, what is at stake is the suggestion
first made by Mishra et al.~\cite{Mishra:2008dm} and further discussed by Mocsy and 
Sorensen \cite{Mocsy:2010um} that the measurement of the harmonic  flow
coefficients $v_n$ for all values of $n$ may provide information about the initial state similar
to the power spectrum extracted from Cosmic microwave background (CMB) radiation.
CMB analysis tools exploit the fact that Hubble expansion dampens vorticity fluctuations,
so that the fluid dynamical evolution stays at all time scales in a linear non-turbulent regime. It is
a priori unclear whether the same situation persists for small fluctuations in  a Bjorken-type expansion, 
or whether small fluctuations can become seeds of turbulent behavior. Our discussion shall address
this question and characterize the limitations of a linear treatment
of fluid dynamic fluctuations in heavy ion collisions, thus gaining some insight into the conditions for
onset of turbulent behavior~\footnote{ 
We remark in this context that turbulence may play a role for heavy ion collisions not only in the more narrow sense of fluid dynamics as we discuss it in this paper. It has been argued that non-Abelian gauge theories show turbulent phenomena such as energy cascades in connection with plasma instabilities and that this could play an important role for the time evolution in the early stage of a heavy-ion collision before a hydrodynamical description which is based on local thermal equilibrium becomes 
applicable~\cite{Arnold:2005qs,Mueller:2006up,Berges:2008mr,Ipp:2010uy}.}.
   
Our work is organized as follows. In section~\ref{sec2}, we show that mild extension of models of 
fluctuating initial conditions give rise to fluctuations in all fluid fields. We note in particular that fluctuations
in the velocity field can have in general a vorticity component  as well as a divergent component, and that
both components may be of comparable size. Furnished with this example that fluctuations in all fluid 
fields may be relevant, we formulate then the equations of motion for fluctuations around a Bjorken
flow field in section~\ref{sec3}, and we solve them in a linearized approximation of the evolution in 
section~\ref{sec4}. We then turn in  section~\ref{sec5} to the case of turbulent fluctuations, when 
non-linear contributions to the equations of motion matter. In particular, we provide a parametric
argument that a large class of fluctuating initial conditions around Bjorken flow evolves at late times
towards an effectively two-dimensional, turbulent system. Motivated by this observation, we recall 
in section~\ref{sec6} pertinent features of turbulence in terms of correlation functions of fluid fields.
In section~\ref{sec7}, we finally relate these remarks to heavy ion phenomenology by showing how
correlations of fluid dynamic fields enter the one- and identical two-particle correlation functions
in a blast wave model supplemented with fluctuations. In the conclusion, we finally summarize
our main findings and provide a short outlook.

\section{Fluctuating initial conditions and vorticity}
\label{sec2}

\begin{figure}[h]
\begin{center}
\includegraphics[width=7.0cm, angle=0]{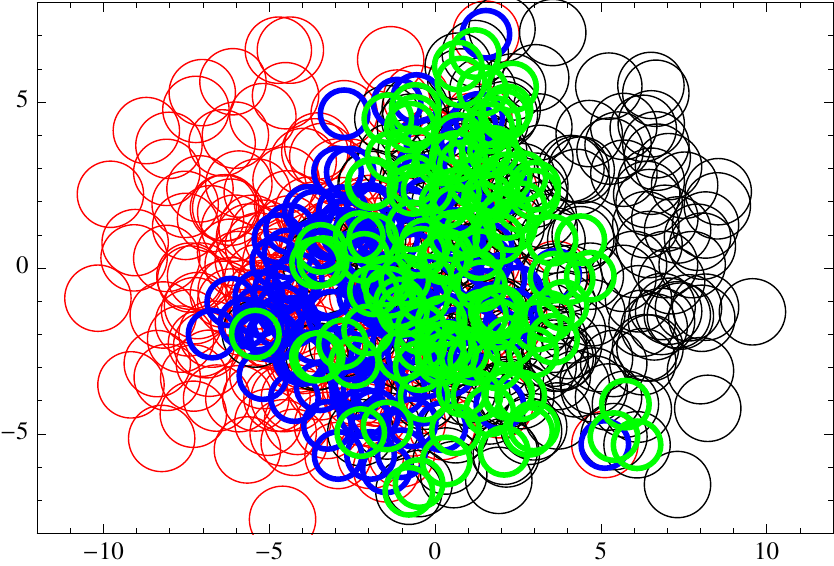}
\includegraphics[width=7.0cm, angle=0]{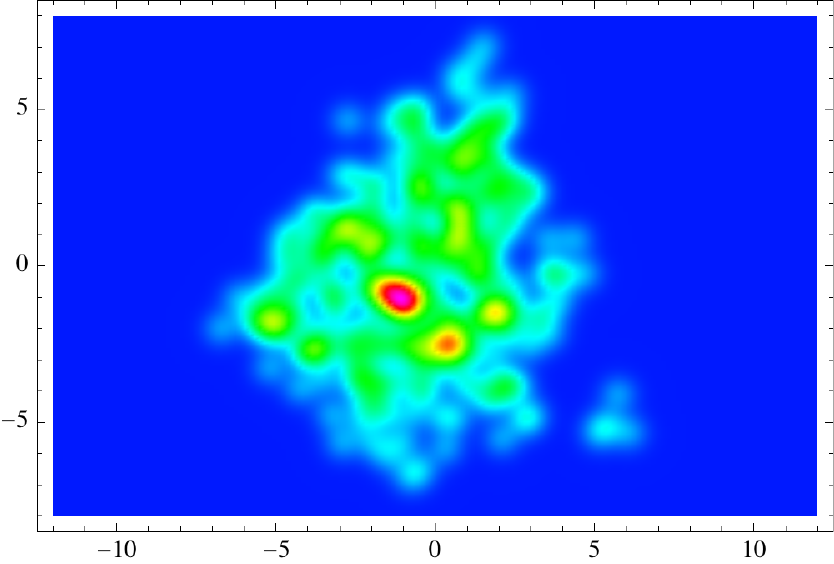}
\caption{(Left hand side) The transverse spatial distribution of nucleons as obtained from a Monte
Carlo Glauber simulation of a Pb-Pb collision at 6 fm  impact parameter. Nucleons of the two colliding 
nuclei are characterized by red and black circles, respectively. The radii correspond to a black 
sphere inelastic nucleon-nucleon cross section of 63 mb at $\sqrt{s_{\rm NN}}= 2.76$ TeV.
'Wounded' nucleons that interact are denoted by smaller 
blue and green circles, respectively. 
(Right hand side) Initial distribution of transverse energy density (\ref{eq2.1}) corresponding to the 
distribution of wounded nucleons on the left hand side. This plot is in arbitrary units. }
\label{fig1}
\end{center}
\end{figure}

We start our discussion with  the prototype of an initial density inhomogeneity
implemented in current event-by-event fluid-dynamical simulations, as used e.g. in Ref.~\cite{Holopainen:2010gz}.
Fluctuations in the initial spatial distribution are described at some initial time $\tau_0$ and
close to mid-rapidity $y=0$ by a 
two-dimensional transverse energy density profile of the form
\begin{equation}
	\epsilon({\bf x}) = \frac{K}{2\pi\sigma^2} \sum_{i=1}^{N_{\rm part}}
		\exp\left[- \frac{\left({\bf x}-{\bf x}_i \right)^2}{2\sigma^2} \right]\, .
		\label{eq2.1}
\end{equation}
 Here, the coordinates ${\bf x}_i$ denote for one specific heavy ion collision
 the positions of wounded nucleons in the transverse plane, as obtained from
a Monte Carlo Glauber simulation, see e.g.~\cite{Miller:2007ri}. A class of events 
corresponds then to a class of independently simulated distributions $\lbrace {\bf x}_i \rbrace$, 
each defining an energy
density $\epsilon({\bf x})$ with different, event-specific fluctuations and each setting the initial
data for an individual fluid dynamic evolution. The normalization $K$ in  (\ref{eq2.1})
can be constrained by data on the total transverse energy produced in the collision 
per unit rapidity~\cite{Holopainen:2010gz}. The smearing parameter $\sigma$ is a model-dependent
input that sets the scale of spatial inhomogeneities. 
Fig.~\ref{fig1} illustrates that this model accounts for significant fluctuations in the
transverse distribution of wounded nucleons and their corresponding energy density
(\ref{eq2.1}). In Fig.~\ref{fig1}, we have chosen a smearing parameter $\sigma = 0.4$ fm,
consistent with previous simulations of event-by-event fluid dynamics~\cite{Holopainen:2010gz}.

One possibility is to initialize fluid dynamical fields at initial time $\tau_0$
with event-wise fluctuations in the transverse energy density $\epsilon({\bf x})$, but with an exactly 
vanishing non-fluctuating flow field in the two transverse
directions $1$ and $2$ at initial time $\tau_0$,
\begin{equation}
	u^1({\bf x},y,\tau_0) = u^2({\bf x},y,\tau_0) \equiv 0\, ,
	\label{eq2.2}
\end{equation} 
and in the rapidity component of the flow vector 
\begin{equation}
	u^y({\bf x},y,\tau_0)  \equiv 0\, .
	\label{eq2.3}
\end{equation} 
However, a larger class of initial conditions is conceivable, since the initial state may also 
display 
fluctuations in the initial flow fields, and since both energy density (\ref{eq2.1})
and the flow fields could depend on rapidity $y$. Fluctuations in the initial velocity fields have
been discussed recently e.g. within a model of the early (Non-Equilibrium) dynamics based on free streaming \cite{Qin:2010pf}. For more discussions of initial flow and its influence on HBT radii see also Refs.~\cite{Pratt:2008qv,Broniowski:2008vp,Jas:2007rw,Broniowski:2008qk,Sinyukov:2011mw}.

{\it Some} fluctuations in the flow field will be generated by the fluid dynamic response to initial
density fluctuations and may therefore be regarded as being implicitly included in the
ansatz (\ref{eq2.1}) - (\ref{eq2.3}).  However,  
such fluid dynamically generated flow fluctuations will be constraint in scale and size
to the fluctuations in energy density, and they will lack by construction some 
qualitative features of the most general fluctuating flow field, such as vorticity.
Vorticity~\footnote{Here, we deviate from the standard notation of vorticity in terms of a cartesian three-vector 
by adopting the notation of vorticity to light cone coordinates, $u^j=(u^1,u^2,u^y)$. We note that the three components of vorticity $(\omega_1,\omega_2,\omega_3)$ do {\it not} form the spatial part of a four-vector and it makes no sense to contract them with the spatial part the metric.}
characterizes the solenoidal part of a general three-dimensional flow field $u^j(x)$, 
\begin{equation}
	\omega_j = (\text{Curl} \; u)_j\, \equiv
\begin{pmatrix}
\frac{1}{\tau}\left(\partial_2 u_y-\partial_y u_2\right) \\ \frac{1}{\tau}\left(\partial_y u_1-\partial_1 u_y\right) \\ \partial_1 u_2 - \partial_2 u_1 \end{pmatrix}.
	\label{eq2.4}
\end{equation}
We note that  the fluid dynamical evolution of fluctuations in 
vorticity and energy density decouples as long as these fluctuations are small and can be treated in a
linearized evolution (see section \ref{sec4}). This is in marked contrast to fluctuations in the irrotational 
part of $u^j$ (sound modes) that can be driven by fluctuations in energy density. 
This indicates that one cannot expect to generate sizable values of vorticity by evolving initial conditions 
of the form (\ref{eq2.1})- (\ref{eq2.3}). However, if fluctuations in vorticity are part of the initial conditions,
then they will propagate and may display particularly interesting dynamical features, as discussed
in sections~\ref{sec4} and ~\ref{sec6}, respectively. 

\begin{figure}[t]
\begin{center}
\includegraphics[height=8.0cm, angle=0]{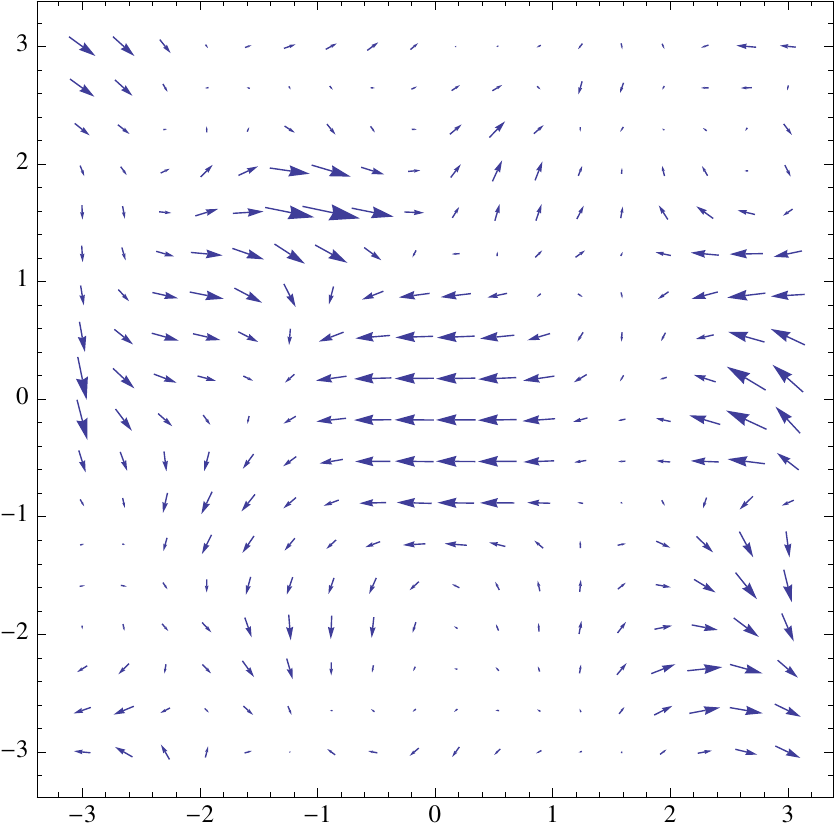}
\caption{Velocity distribution corresponding to Fig.~\ref{fig1} if one assumes for the contribution of each wounded nucleon a random velocity in the transverse plane drawn from a Gaussian distribution.
Shown is the innermost area $-3.5\text{ fm}< x^1, x^2 < 3.5\text{ fm}$ of the transverse plane.}
\label{fig2}
\end{center}
\end{figure}

To set the stage for our discussion in later section, we demonstrate now that relatively mild extensions 
of (\ref{eq2.1})
can lead naturally to fluctuations in velocity, including a non-vanishing solenoidal component
(\ref{eq2.4}). It is one arguably mild extension of (\ref{eq2.1}) to associate the transverse region
around a single wounded nucleon not with an energy density, but with an energy-momentum
tensor $T^{\mu\nu}_w$, such that the initial energy-momentum tensor of the entire nucleus-nucleus
collision takes the form
\begin{equation}
	T^{\mu\nu}(\tau, x^1, x^2,y)=
	\sum_{i=1}^{N_{\rm part}} T^{\mu\nu}_w (\tau, x^1-x_i^1, x^2-x_i^2,y) \, .	
	\label{eq2.5}
\end{equation}
In general, equation (\ref{eq2.5}) can account for non-vanishing fluctuating initial conditions
in both energy density $\epsilon(x)$ and flow $u^j(x)$. For instance, neglecting for simplicity
non-ideal, shear viscous contributions to (\ref{eq2.5}) and
assuming an ideal equation of state $\epsilon(x) = 3\, p(x)$, one can write the 
initial conditions for $\epsilon$ and $u^j$ at some fixed rapidity $y$ and initial time $\tau_0$ to linear order in $u^j$ in the form
\begin{equation}
	\epsilon(x^1,x^2) \left(1,\, \frac{4}{3}u^j(x) \right) \equiv T^{\mu 0}(\tau_0,x^1,x^2,y).
	\label{eq2.6}
\end{equation}
The transverse energy density associated to a single wounded nucleon is given by 
$\epsilon_w({\bf x})= T^{00}_w({\bf x})$. Therefore, the $0$-component of (\ref{eq2.6})
defines an equation of the type (\ref{eq2.1}) for the energy density,
\begin{equation}
	\epsilon(x^1,x^2) = \sum_{i=1}^{N_{\rm part}} \epsilon_w (x^1-x_i^1, x^2-x_i^2)\, ,
	\label{eq2.7}
\end{equation}
but it has also 
a fluctuating initial flow field defined by the spatial components of (\ref{eq2.6}),
\begin{equation}
	u^j({\bf x}) = \frac{ \sum_{i=1}^{N_{\rm part}} \epsilon_w (x^1-x_i^1, x^2-x_i^2)
	\, u^j_w(x^1-x_i^1, x^2-x_i^2)}{ \sum_{i=1}^{N_{\rm part}} \epsilon_w (x^1-x_i^1, x^2-x_i^2)}\, .
	 \label{eq2.8}
\end{equation}
Here $u^j_w$ is defined by writing equation (\ref{eq2.6}) for the energy momentum tensor
associated to a single wounded nucleon.  The size of initial fluctuations in the velocity field
(\ref{eq2.8}) depends on how the initial transverse motion associated to the 
the single wounded nucleon is modeled. Taking guidance from blast wave models, one
may choose for $u^j_w$ e.g.~an azimuthally symmetric radial flow field with some radial dependence $w$, 
$u^j_w({\bf x}) = \frac{x^j-x^j_i}{\vert {\bf x}-{\bf x}_i\vert}\, w(\vert {\bf x}-{\bf x}_i\vert)$ for $j=1,2$ and $u^y=0$,
say. For such an ansatz, one checks easily that (\ref{eq2.8}) defines in general 
a flow field of non-vanishing transverse divergence and non-vanishing longitudinal vorticity,
\begin{equation}
 \begin{split}
  \partial_1\, u^1({\bf x}) + \partial_2\, u^2({\bf x}) &\not= 0\, ,\\
 \omega_3({\bf x})  &\not= 0\, .
 \end{split}
 \label{eq2.9}
\end{equation}
A more general ansatz may be based on the observation that in general, 
the transverse energy deposited by a single wounded nucleon in a finite window of
rapidity recoils against transverse momentum outside this rapidity window. This 
argues for a net transverse velocity component ${\bf v}_i$ associated to the contribution of each
wounded nucleon in (\ref{eq2.5}).  To illustrate this effect, we assume that each
wounded nucleon in the sample shown in Fig.~\ref{fig1} is associated with a non-relativistically
small random transverse velocity component ${\bf v}_i$, drawn from a Gaussian distribution of
width $\langle \vert {\bf v}\vert \rangle = 0.1 c$~\footnote{We note that the value of the width does not play any role for the subsequent arguments and could easily be smaller in a realistic situation.}. The resulting initial transverse flow field (\ref{eq2.8})
is shown in Fig.~\ref{fig2}. By comparing to Fig.~14 of Ref.~\cite{Qin:2010pf}, we note that a seemingly
comparable flow field can be obtained in a model of early non-equilibrium dynamics based on
free streaming where the contribution of every particle is taken to be delocalized in position space 
over a small volume. 

\begin{figure}[t]
\begin{center}
\setlength{\unitlength}{\textwidth}
\begin{picture}(1,0.4)
\put(0.05,0){\includegraphics[width=7.0cm, angle=0]{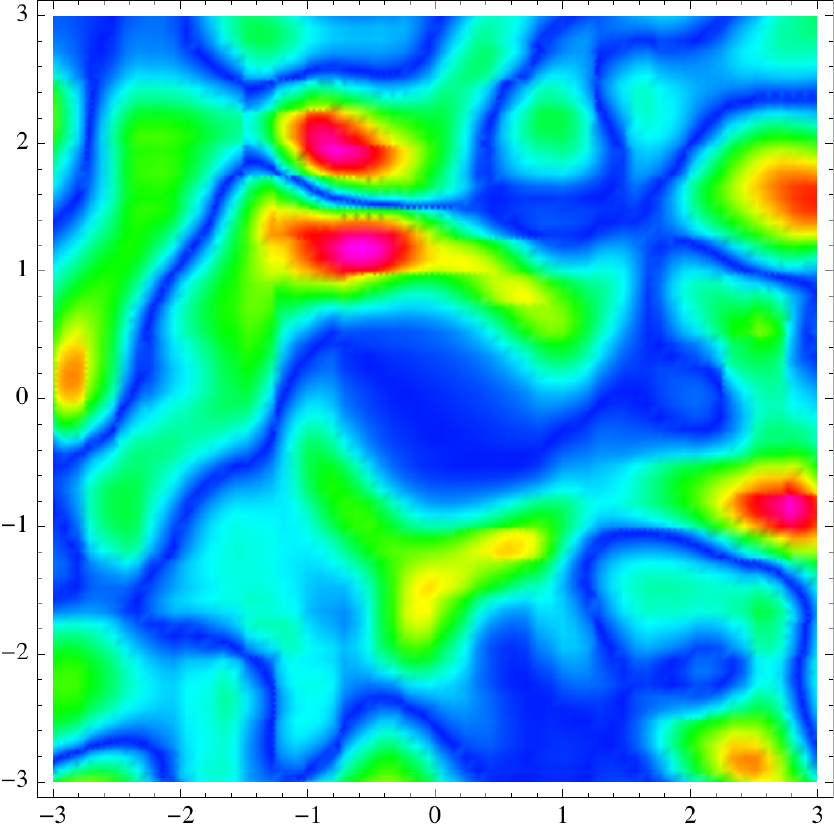}}
\put(0.45,0){\includegraphics[width=7.0cm, angle=0]{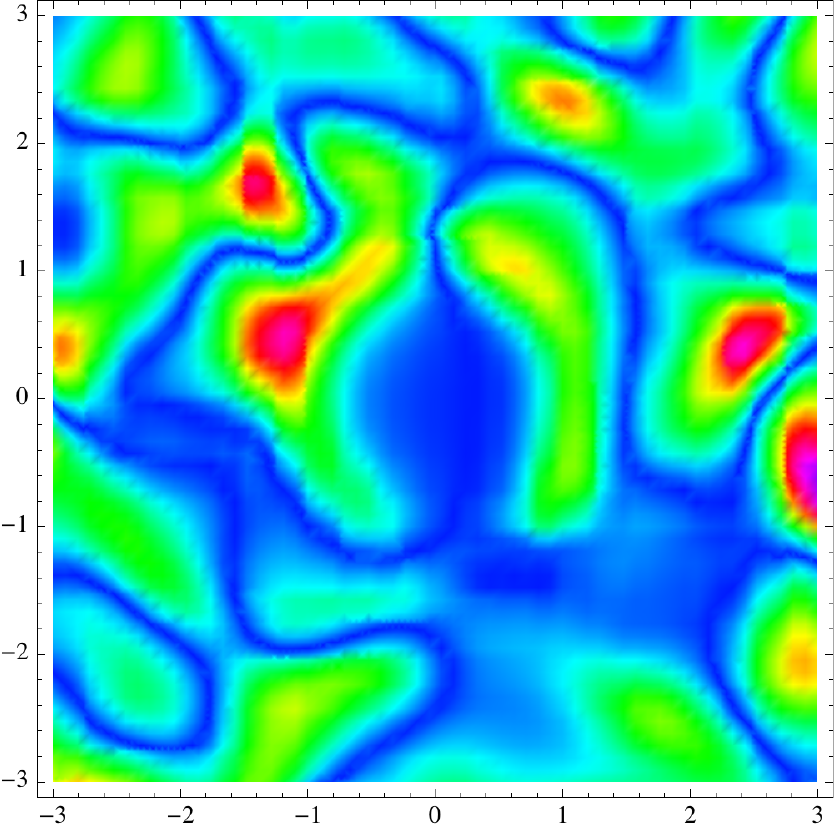}}
\put(0.86,0.1){\includegraphics[width=1.5cm, angle=0]{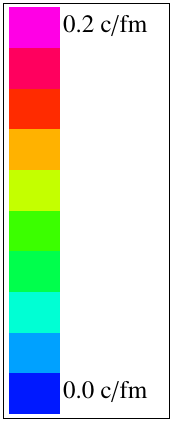}}
\end{picture}
\caption{The left hand side shows the absolute value of vorticity $\vert \partial_1 u^2-\partial_2 u^1\vert$ for the velocity field shown in Fig. \ref{fig2}. Similarly, the right hand side shows the absolute value of the divergence of the fluid velocity $\vert \partial_1 u^1+\partial_2 u^2 \vert$ for the same velocity field. The color coding is the same on both sides.}
\label{fig2b}
\end{center}
\end{figure}

In Fig.~\ref{fig2b}, we plot the absolute value of the longitudinal vorticity $\vert \omega_3 \vert = \vert \partial_1 u^2-\partial_2 u^1\vert$ and the transverse divergence $\vert \partial_1 u^1+\partial_2 u^2 \vert$ for the flow field
shown in Fig.~\ref{fig2}.  Inspection of this figure shows that both components fluctuate with a similar magnitude and over similar transverse length scales.

For illustrative purposes, we have chosen in Fig.\ \ref{fig2b} velocity fluctuations of an average strength $\langle | {\bf v} | \rangle = 0.1$. We emphasize, however, that none of our conclusions in this section or in the following sections depends on the precise numerical choice for $\langle | {\bf v} | \rangle $. In particular, repeating the analysis of Fig.\ \ref{fig2b} for much smaller values of $\langle |{\bf v }| \rangle$ would equally well support the only conclusion that we draw from it, namely that the irrotational and solenoidal components of the velocity may be of comparable size. We observe at this point that there is no general model-independent argument for the relative size fluctuations in $u^j$ and $\epsilon$. This motivates us in the following sections to treat {\it all} conceivable sources of event-by-event fluctuations on an equal footing.

\section{Fluid dynamic equations of motion for relativistic heavy ion collisions}
\label{sec3}
On time and length scales that are large compared to the typical relaxation times and lengths for
thermal and chemical equilibration, relativistic fluid dynamics provides an effective
theory for the multi-particle system produced in heavy ion collisions. A large set of
experimental observations support the assumption that in ultra-relativistic heavy
ion collisions the range of validity of this effective fluid dynamical description is large and 
comprises bulk hadron production up to a few GeV in transverse 
momentum~\cite{Heinz:2009xj,Romatschke:2009im,Teaney:2009qa}.
In this section, we recall first shortly the fluid dynamic equations of motion. We focus then on the
Bjorken model that defines a particularly simple expanding geometry and  that encodes important 
features of a relativistic heavy ion collision. Regarding the Bjorken model as defining the 
average fluid dynamical field, we then discuss how
fluctuations in energy density and flow propagate in the expanding geometry of a relativistic
heavy ion collision. 

The relativistic hydrodynamic equations for a fluid without any conserved charges read~\cite{Heinz:2009xj,Romatschke:2009im,Teaney:2009qa} 
\begin{equation}
\begin{split}
D \epsilon + (\epsilon+p) \partial_{\mu} u^{\mu} + \pi^{\mu\nu} \Delta_{\mu}^{\;\;\alpha} \partial_{\alpha} u_{\nu} & = 0,\\
(\epsilon + p) D u^{\alpha} + \Delta^{\alpha\beta} \partial_{\beta} p - \Delta^{\alpha}_{\;\;\nu} \partial_{\mu} \pi^{\mu\nu} & = 0.
\end{split}
\label{eq3.1}
\end{equation}
Here $\epsilon$ is the energy density and $p$ is the pressure in the fluid rest frame, $\pi^{\mu\nu}$ is the viscous part of the energy-momentum tensor in the Landau frame, $u_{\mu} \pi^{\mu\nu} = 0$. The partial derivative $\partial_\alpha$ must be replaced by the covariant derivative $\nabla_{\alpha}$ if one works with coordinates other than cartesian. We work with a cartesian metric of signature $g^{\mu\nu}=\text{diag}(-1,1,1,1)$. The matrix $\Delta^{\mu\nu}$ projects to the subspace orthogonal to the fluid velocity, $\Delta^{\mu\nu} = g^{\mu\nu} + u^{\mu} u^{\nu}$. The derivative in the direction of the fluid motion is $D=u^\mu \partial_\mu$.

The viscous part of the energy-momentum tensor can be expanded in a derivative expansion. To lowest order it vanishes, leading to ideal hydrodynamics. The first order contains shear and bulk viscocity terms
\begin{equation}
\pi^{\mu\nu} = - 2 \eta \sigma^{\mu\nu} + \zeta \Delta^{\mu\nu} \nabla_{\alpha} u^{\alpha},
\end{equation}
where
\begin{equation}
\sigma^{\mu\nu} = \frac{1}{2}(\Delta^{\mu}_{\;\;\alpha} \nabla^{\alpha} u^{\nu} + \Delta^{\nu}_{\;\;\alpha} \nabla^{\alpha} u^{\mu})-\frac{1}{3}\Delta^{\mu\nu}(\nabla_{\alpha}u^{\alpha})
\end{equation}
is transverse (orthogonal to $u^{\mu}$) and traceless.

To second order in the gradient expansion, the fluid dynamic equations of motion contain various relaxation time corrections. It is a peculiar feature of a second order approximation that the evolution equations are
hyperbolic and propagation is limited to the forward light cone even for perturbations of large
wave-vector $k$. For this reason, second order fluid dynamics is often referred to as causal viscous
fluid dynamics. However, this wanted feature of causality is not guaranteed to persist in higher orders of 
the gradient expansion. More generally,  fluid dynamics is a long distance effective theory that by its very
construction cannot be expected to be reliable for large wave-vectors. For the propagation of fluctuations with
small gradients (i.e. small wave-vectors $k$), second order fluid dynamics will make only small corrections 
to a first order treatment.  For this reasons and to keep the formalism simple, we restrict the discussion in the present paper to first order fluid dynamics. For the case of vanishing bulk viscocity $\zeta$, the corresponding equations
of motion read
\begin{equation}
D \epsilon + (\epsilon + p) \nabla_{\mu} u^\mu - 2 \,\eta \,\sigma_{\mu\nu} \sigma^{\mu\nu} = 0,
\label{eq:firstorderhyrdoepsilon}
\end{equation}
\begin{equation}
(\epsilon+p) D u^{\nu} + \Delta^{\nu\mu} \nabla_{\mu} p - 2 \,\eta\, \Delta^{\nu}_{\;\;\alpha}\nabla_{\mu} \sigma^{\mu\alpha} = 0\, .
\label{eq:firstorderhydrou}
\end{equation}
There is evidence that dissipation in a heavy ion collision is mainly due to shear viscosity, and we therefore neglect bulk viscosity for the remainder of this paper.
\subsection{The Bjorken model}
\label{sec3a}
We are interested in studying fluid dynamic fluctuations in the expanding geometry of a relativistic heavy
ion collision. The Bjorken model is arguably the simplest formulation of a corresponding expanding
geometry. Motivated by the idea that in nuclear collisions at ultra-relativistic energy, particle production 
is almost flat in rapidity, Bjorken~\cite{Bjorken:1982qr} proposed to formulate initial conditions for fluid dynamic
fields that are independent of space-time rapidity $y=\text{arctanh}(x_3/|x_0|)$, that means
$\epsilon(\tau,{\bf x},y) = \epsilon(\tau,{\bf x})$ and 
$u^\mu = \left(\sqrt{1+{\bf u}^2}, {\bf u}, u^y\right) = \left(\sqrt{1+{\bf u}^2}, {\bf u}, 0\right)$. If this
condition is satisfied at some initial proper time $\tau = \tau_0$, then it persists for all proper times
$\tau=\sqrt{x_0^{2}-x_3^{2}}$ throughout the evolution. This renders the longitudinal evolution trivial,
and the numerical task simplifies to the solution of a (2+1)-dimensional problem.

Under the further simplifying assumption that the initial transverse flow field vanishes at initial
times and that transverse gradients in energy density are absent,  the Bjorken model reduces 
to an effectively (1+1)-dimensional toy model that allows for an explicit analytical treatment.
In this case, the evolution equation 
for the energy density becomes
\begin{equation}
\partial_{\tau} \epsilon + \frac{\epsilon+p}{\tau} - \eta \frac{4}{3 \tau^{2}} = 0\, ,
\label{eq:evolutioneqepsilon}
\end{equation}
where energy density and pressure are related by the equation of state $\epsilon = \epsilon(p)$. 
For what follows, it will be useful to rewrite this equation in terms of the enthalpy 
\begin{equation}
w=\epsilon+p = sT\, ,
\end{equation}
and the kinematic viscosity 
\begin{equation}
  \nu = \eta/w\, .
\end{equation} 
Eq.\ \eqref{eq:evolutioneqepsilon} reads then
\begin{equation}
\partial_{\tau} \epsilon + \frac{w}{\tau}\left(1-\frac{4\nu}{3\tau}\right)= 0\, .
\label{eq:evolutioneps2}
\end{equation}
Throughout this work, we shall neglect terms that are parametrically suppressed by
powers of $\nu/\tau$ compared to some other term of the same structure. With this
approximation, equation \eqref{eq:evolutioneps2} becomes independent of shear viscosity. As will become clear in the following, however, the dominant effect of shear viscosity on fluctuations can be retained within this approximation. 

With the approximation $\nu/\tau \ll 1$
and for an ideal equation of state $\epsilon = 3 p$, one
then finds the characteristic time dependencies of the Bjorken model for  energy density
\begin{equation}
\epsilon_\text{Bj}(\tau) = \epsilon_\text{Bj}(\tau_0) \left(\frac{\tau_0}{\tau}\right)^{4/3}\, ,
\label{eq:BjorkenEnergydensity}
\end{equation}
and temperature 
\begin{equation}
T_\text{Bj}(\tau) = T_\text{Bj}(\tau_0) \left(\frac{\tau_0}{\tau}\right)^{1/3}\, .
 \label{eq3.11}
\end{equation}
For a time-independent normalized viscosity $\eta/s$, the ratio 
\begin{equation}
\frac{\nu_\text{Bj}(\tau)}{\tau} = \frac{\nu_\text{Bj}(\tau_0)}{\tau_0} \left(\frac{\tau_0}{\tau}\right)^{2/3}
\label{eq:Bjorkennutau}
\end{equation}
decreases. Therefore, replacing the bracket in Eq.\ \eqref{eq:evolutioneps2} by unity is an
approximation that is consistent with the late time behavior.

We note at this point that in situations with strong (non-Gaussian) fluctuations, the evolution equation for averaged fields such as energy density $\epsilon(\tau) = \langle \epsilon(\tau) \rangle$ gets modified by additional terms, see the discussion in Sect.\ \ref{sec6.1}. For the present paper we assume that these modifications are small and can be neglected.

\subsection{Fluctuations on top of a Bjorken background field}
\label{sec3b}

In this section, we formulate the theory of the dynamics of fluctuations on top of a Bjorken background
field without transverse gradients. That means that the hydrodynamical fields $u^\mu$, $\epsilon$ 
when averaged over many events follow a Bjorken type solution. 
However, \emph{locally} and for a \emph{particular event} we expect deviations which we want to investigate in more detail.
We have chosen a Bjorken background field for our study mainly
for two reasons. First, the analytical simplicity of this background will allow for a particularly explicit
discussion. Second,  the Bjorken model contains essential features of realistic expansion scenarios of
relativistic heavy ion collisions.~\footnote{
It has been pointed out repeatedly that despite its simplicity, the Bjorken model without transverse 
gradients retains important features of the early time dynamics of heavy ion 
collisions. The argument is based on the observation that event-averaged initial  energy density 
distributions show typically only small transverse gradients in the central region of the transverse
plane;  the central region may indeed be approximated by the ansatz ${\bf u} = 0$, 
$\epsilon({\bf x}) = \epsilon$. The transverse evolution of this initial condition may then be thought 
of qualitatively as being dominated by a rarefaction wave that moves from the outside (vacuum) to more 
and more central positions in the transverse plane at late times. At a given position in the transverse
plane, the dynamical evolution may be viewed as being characterized by the effectively (1+1)-
dimensional Bjorken model up to the later time at which the rarefaction wave 
reaches the corresponding transverse position. Based on such considerations, the Bjorken
estimates for the time-dependence of energy-density (\ref{eq:BjorkenEnergydensity}) and 
temperature (\ref{eq3.11}) are used regularily in simple phenomenological estimates.}

We denote fluctuations on top of the Bjorken flow $u_{\rm Bj}^{\mu} = \left(1,0,0,0 \right)$ by relaxing the 
constraints (\ref{eq2.2}) and (\ref{eq2.3}) and allowing for local fluctuations in the transverse
and rapidity components, $u^1, u^2, u^y$. The normalization condition $u^\mu u_\mu = -1$ 
of the local fluid velocity $u^\mu=(u^\tau, u^1,u^2,u^y)$ implies then 
\begin{equation}
(u^\tau)^2 = 1 + (u^1)^2+ (u^2)^2 + \tau^2 (u^y)^2 = 1 + u_j u^j.
\label{eq:normalizationutau}
\end{equation}
Here and in what follows, we work in light-cone coordinates $\tau, x_1, x_2, y$ with metric 
$g^{\mu\nu} =\text{diag}(-1,1,$ $1,1/\tau^2)$. The latin index $j$ is summed over $1,2,y$ and 
the corresponding three-dimensional metric reads  $g^{ij} = \text{diag}(1,1,1/\tau^2)$. 
We consider small local fluctuations in the sense that $u_ju^j(x) \ll 1$.

In the following we neglect terms that are parametrically suppressed due to $u_j u^j(x) \ll 1$ or due to $\nu/\tau \ll 1$ compared to other terms with the same combination of derivatives of the velocity and pressure fields. We note that for every combination of derivatives there is one term of lowest order which is {\it not} neglected and that the main physical effects of viscosity -- the damping of velocity fluctuations and the dissipation of kinetic energy to heat -- are correctly taken into account. With this approximation scheme we find from Eq.\ \eqref{eq:firstorderhyrdoepsilon} and \eqref{eq:firstorderhydrou} the following equations governing the velocities in the transverse plane ($j=1,2$) and in rapidity direction ($j=y$)
\begin{equation}
\partial_\tau u_j+ u^i \partial_i u_j + \frac{1}{w} \left[ \partial_j p + u_j (\partial_\tau p + u^i \partial_i p) \right] 
- \nu \left[ \frac{1}{3} \partial_j \partial_i u^i + \partial_i \partial^i u_j \right] = 0\, .
\label{eq:A}
\end{equation}
Here, the first two terms describe the change in the velocity along the direction of the fluid motion. They
result from writing $D u_j = u^\mu \partial_\mu u_j$ for small deviations from  the Bjorken background. 
The terms in the first square bracket account for two effects. One is the acceleration of the fluid due to pressure gradients in the transverse direction. The second term proportional to $u_j$ is dominated by the decrease of pressure for increasing $\tau$. This dilution of the fluid leads to an acceleration in the direction of $u_j$. Finally, 
there are effects of viscosity that are similar to the corresponding term in the (non-relativistic) Navier-Stokes equation. 

In addition to eq.(\ref{eq:A}), one finds under the same assumptions for small local fluctuations around 
the Bjorken background the equation of motion for the internal energy density
\begin{equation}
\partial_\tau \epsilon + u^j \partial_j \epsilon + w \left[\frac{1}{\tau} + \partial_j u^j \right]
- \eta \left[ \partial_i u_j \partial^i u^j + \partial_i u_j \partial^j u^i - \frac{1}{3} \partial_i u^i \partial_j u^j\right] = 0.
\label{eq:C}
\end{equation}
 Here, the first two terms describe the change along the fluid direction of motion. The first square bracket describes dilution effects; the first term $\sim \frac{1}{\tau}$ is due to the expansion of the Bjorken-background in the longitudinal direction while the second term measures the effect of a possible dilution (or compression) in the transverse and rapidity directions. Viscous correction that are parametrically suppressed due to $\eta/ (w\tau)\ll 1$ have been dropped, and the remaining dissipative contribution to the evolution of internal energy are given
 in the last bracket of (\ref{eq:C}). They describe how kinetic energy is transferred from the macroscopic motion of the fluid to internal energy.

It will turn out to be useful to rewrite Eqs.\ \eqref{eq:A} and \eqref{eq:C} in terms of rescaled fluctuations in
velocity, 
\begin{equation}
u_j = \left(\frac{\tau}{\tau_0}\right)^{1/3} v_j\, ,
\end{equation}
and for a rescaled time variable
\begin{equation}
t = \frac{3 \,\tau^{4/3}}{4\, \tau_0^{1/3}}, \quad \partial_t \equiv \left(\frac{\tau_0}{\tau}\right)^{1/3} \partial_\tau\, .
\label{eq:deft}
\end{equation}
(Of course, this rescaled time $t$ is {\it not} the time variable $x^0$ in the laboratory frame.)
In what follows, we also absorb deviations from the $\tau$-dependence of Bjorken's 
energy density (\ref{eq:BjorkenEnergydensity}) in terms of the quantity
\begin{equation}
d_{\tau}\equiv \left(\frac{\tau_0}{\tau}\right)^{2/3}\, d\, ,
\qquad d \equiv \ln\left[\frac{T}{T_\text{Bj}(\tau)}\right].
\label{eq3.18}
\end{equation}
Finally, we assume that the shear viscosity $\nu$ follows the Bjorken behavior \eqref{eq:Bjorkennutau}. That means, we neglect local fluctuations in the kinematic viscosity since they are expected to have only a minor effect. The kinematic viscosity $\nu_0$ can then be written as
\begin{equation}
\nu_0 = \nu \left(\frac{\tau_0}{\tau}\right)^{1/3} = \nu_\text{Bj}(\tau_0).
\label{eq:defnu0}
\end{equation}

For an ideal equation of state $\epsilon=3p$, using  $\tfrac{1}{w} dp=\tfrac{1}{sT}\tfrac{\partial p}{\partial T} dT = \tfrac{1}{T} dT$, and neglecting a dissipation term $\sim \nu$ that is of higher power in the velocity field, one can show that Eqs.\ \eqref{eq:A} - \eqref{eq:C} lead to the equation for the (rescaled) velocity ($j=1,2,y$)
\begin{equation}
 \partial_t v_j + \sum_{m=1}^2 v_m \partial_m v_j + \frac{1}{\tau^2} v_y \partial_y v_j + \partial_j d_{\tau} -\frac{1}{3} \vartheta \,v_j,
 -\nu_0 \left[ \frac{1}{3} \partial_j \vartheta + (\partial_1^2+\partial_2^2+\frac{1}{\tau^2} \partial_y^2) v_j \right] =0
\label{eq:velocityeqtransformed}
\end{equation}
and for the quantity $d_{\tau}$
\begin{equation}
\begin{split}
& \partial_t d_{\tau} + \frac{1}{2 t} d_{\tau} + \sum_{m=1}^2 v_m \partial_m d_{\tau} + \frac{1}{\tau^2} v_y \partial_y d_{\tau} + \frac{1}{3} \left( \frac{\tau_0}{\tau} \right)^{2/3} \vartheta  
 - \frac{\nu_0}{6} {\Bigg [} \sum_{m,n=1}^2 (\partial_m v_n + \partial_n v_m)(\partial_m v_n+\partial_n v_m) \\
&\qquad \qquad \qquad \qquad
 + \frac{2}{\tau^2}\sum_{m=1}^2 (\partial_y v_m+\partial_m v_y)(\partial_y v_m+\partial_m v_y)
 + \frac{4}{\tau^4} (\partial_y v_y)^2 -\frac{2}{3} \vartheta^2 {\Bigg ]}=0.
\end{split}
\label{eq:deqtransformed}
\end{equation}
Here and in what follows, we denote the expansion scalar $\partial_j\, v^j$ of the rescaled velocity
fields by 
\begin{equation}
\vartheta = \partial_1 v_1 +\partial_2 v_2 + \frac{1}{\tau^2} \partial_y v_y.
\label{eq:defvartheta}
\end{equation}
In the following we will use both the representation of the equations of motion in Eq.\ \eqref{eq:A}, \eqref{eq:C} and the one in \eqref{eq:velocityeqtransformed}, \eqref{eq:deqtransformed}. In particular the discussion of linear fluctuations in section \ref{sec4} will be largely based on \eqref{eq:A} and \eqref{eq:C}, while for the discussion of non-linear fluctuations in section \ref{sec5} the representation \eqref{eq:velocityeqtransformed}, \eqref{eq:deqtransformed} will be more appropriate.

\section{Linear fluctuations}
\label{sec4}
In this section we discussion the evolution of fluid dynamical fluctuations that are small enough to neglect non-linear terms in \eqref{eq:A} and \eqref{eq:C}.  In addition, it is assumed that the deviation of the temperature field from the homogeneous background is small, $d\ll 1$. The resulting linearized equations describe laminar flow. They apply to systems with sufficiently small Reynolds number, as we shall discuss in section~\ref{sec5}.

For a fixed time $\tau$ and given spatial boundary conditions, one can divide the velocity field uniquely
into a solenoidal part with vanishing divergence, and an irrotational part with vanishing curl, 
\begin{equation}
	\begin{split}	
  u_j &= u_j^S+u_j^I\, ,\\
\text{Div}\; u^S &\equiv 0\, ,\\
  \text{Curl} \; u^I &\equiv 0\, ,
    \end{split}
    \label{eq4.1}
\end{equation}
where $\text{Div}\; u = \partial_j u^j$ and $\text{Curl}\;u$ is defined in Eq.\ (\ref{eq2.4}). 
We recall that the derivative operators in (\ref{eq4.1}) introduce an explicit $\tau$-dependence since they 
involve the three-dimensional metric $g_{ij} = \text{diag}(1,1, \tau^2)$. Therefore, the splitting of $u_j$
into a solenoidal and an irrotational part does not commute with the $\tau$-derivative. 

The field $u_j^I$ can be represented by the expansion scalar \eqref{eq:defvartheta}
\begin{equation}
\theta =
\underbrace{\partial_1 u_1+\partial_2 u_2}_{\equiv \theta_T }
+
\underbrace{\frac{1}{\tau^2}\partial_y u_y}_{\equiv \theta_y}
=\left(\frac{\tau}{\tau_0}\right)^{1/3} \vartheta,
\end{equation}
 It will be convenient to write this expansion scalar as a sum of a transverse and a longitudinal
 contribution,  $\theta = \theta_T + \theta_y$.

From \eqref{eq:A}, \eqref{eq:C} we obtain the linearized equations
\begin{eqnarray}
\partial_\tau \theta_T - \frac{1}{3\tau} \theta_T + (\partial_1^2+\partial_2^2) d 
- \nu \left[\tfrac{1}{3} (\partial_1^2+\partial_2^2) \theta+(\partial_1^2+\partial_2^2+\tfrac{1}{\tau^2}\partial_y^2) \theta_T \right] &=& 0,\label{eq:hatvartheta_T}\\
\partial_\tau \theta_y + \frac{5}{3\tau} \theta_y + \tfrac{1}{\tau^2}\partial_y^2 d 
- \nu \left[\tfrac{1}{3} \tfrac{1}{\tau^2} \partial_y^2 \theta+(\partial_1^2+\partial_2^2+\tfrac{1}{\tau^2}\partial_y^2) \theta_y \right] &=& 0,\label{eq:hatvartheta_y}\\
\partial_\tau d + \frac{1}{3} \theta &=&0,\label{eq:hatd}\\
\partial_\tau \omega_j -\frac{h_j}{3\tau} \omega_j - \nu (\partial_1^2+\partial_2^2+\tfrac{1}{\tau^2}\partial_y^2) \omega_j &=& 0.
\label{eq:hatomega}
\end{eqnarray}
Here, we recall that the quantity $d$ denotes the logarithmic temperature \eqref{eq3.18}. 
The symbol $h_j$ in \eqref{eq:hatomega} takes the values $h_1=h_2=-2$ and $h_3=1$.

Interestingly, the vorticity modes $\omega_j$ decouple from the velocity divergence $\theta$, the
logarithmic temperature field $d$ and from each other. Using Fourier decomposition with respect to the spatial argument, 
\begin{equation}
 \omega_j(\tau,x_1,x_2,y) =
 \int \frac{d^3k}{(2\pi)^3} \; \omega_j(\tau,k_1,k_2,k_y) \; e^{i (k_1 x_1+k_2 x_2+k_y y)}\, ,
\end{equation}
their diffusion-type equation of motion can be directly solved
\begin{equation}
\omega_j(\tau,k_1,k_2,k_y)  =  \omega_j(\tau_0,k_1,k_2,k_3)  
\times \left(\frac{\tau}{\tau_0}\right)^{h_j/3} e^{-\nu_0 (k_1^2+k_2^2)(t-t_0)+\nu_0 \tfrac{9}{8 \sqrt{t_0}}k_y^2 \left(\tfrac{1}{\sqrt{t}}-\tfrac{1}{\sqrt{t_0}}\right)}.
\label{eq:vorticitywave}
\end{equation}
%
\begin{figure}[t]
\begin{center}
\includegraphics[height=12.0cm, angle=0]{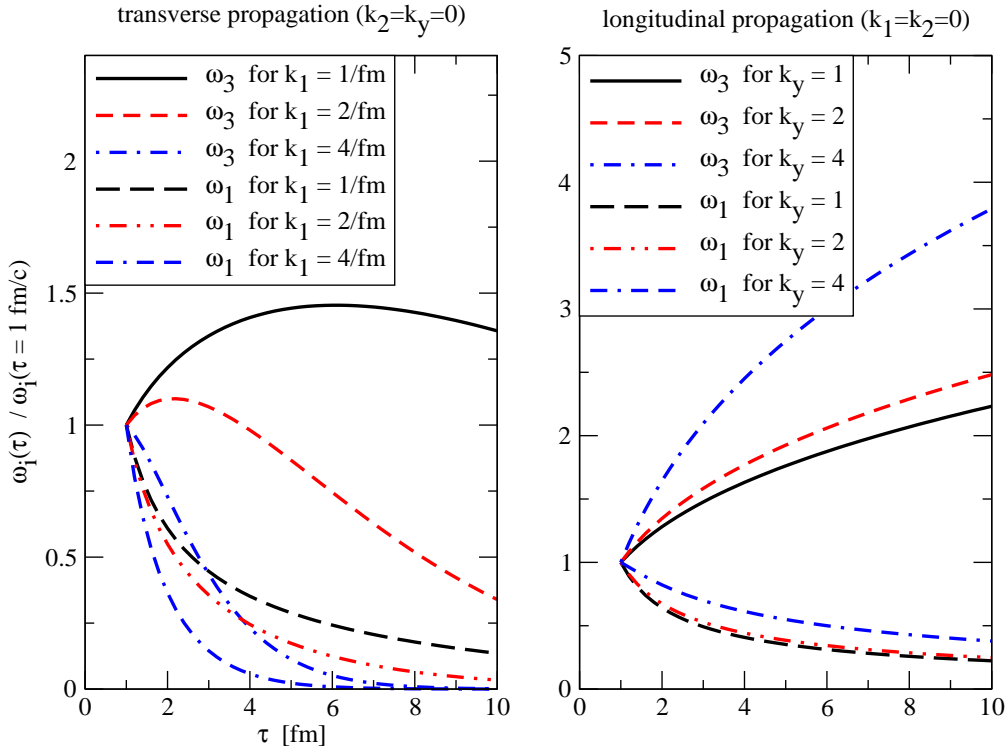}
\caption{
Time dependence of the normalized transverse ($i=1$) and longitudinal ($i=3$) vorticity
amplitude (\protect\ref{eq:vorticitywave}) for modes with wave vectors $k_1$, $k_2=k_y=0$
 in the transverse direction (left hand side) and with wave vectors $k_y$, $k_1=k_2=0$
 in the longitudinal direction (right hand side). Input values are $T(\tau = 1\text{fm/c}) = 500$ MeV
 and $\eta/s = 1/4\pi$. 
}
\label{fig3}
\end{center}
\end{figure}

We assume here a constant $\nu_0$ as defined in \eqref{eq:defnu0} and we use \eqref{eq:deft}.  
One sees from equation (\ref{eq:vorticitywave}) that 
vorticity modes for essentially all wave vectors are dominated at late times
by an exponentially decaying function with
a decay time set by the product of kinematic viscosity and the square of the wave vector. 
A somewhat unusal case is the time evolution 
of modes with $k_y\neq 0$ where the exponential damping term is modified by a term 
$\propto 1/\sqrt{t}$.  In particular, for $k_1^2+k_2^2=0$ but $k_y\neq 0$ the vorticities
do not decay exponentially for $\tau\to\infty$. In addition, the exponential decay is modified by a term 
that decreases algebraically for the transverse components $\omega_1$ and $\omega_2$ 
and that increases in the longitudinal components $\omega_3$. For finite times, the 
algebraic increase of $\omega_3$ can overcome the exponential decay with viscosity.

In Fig.~\ref{fig3}, we plot the solution (\ref{eq:vorticitywave}) of the linearized fluid dynamic equations 
of motion for phenomenologically motivated input values, namely a small normalized shear
viscosity $\eta/s = 1/4\pi$ and a temperature of $500$ MeV at initial time $\tau_0=1$ fm/c. This translates
into an initial kinematic viscosity $\nu_0 \simeq 0.03$ fm. Most generally, Fig.~\ref{fig3} illustrates
the interplay between an exponential decay set by kinematic viscosity, and the characteristic
algebraic dependencies of the transverse and longitudinal vorticity components. More
specifically, we have chosen in Fig.~\ref{fig3} wave vectors that correspond to fluctuations on length 
scales between 1 fm and 0.25 fm, as may be regarded as realistic for the initial state of 
the system created in heavy ion collisions. We observe from Fig.~\ref{fig3} that such fluctuations
are modified but persist over time scales of $O(10\, \text{fm/c})$ typical for the expansion history
of relativistic heavy ion collisions. The figure illustrates that over times scales relevant for the fluid
dynamic expansion of heavy ion collisions, some fluctuations on phenomenologically relevant
scales may get amplified rather than dampened. 
Moreover, the relative attenuations (or amplifications) of
vorticity components over times of $O(10\, \text{fm/c})$ are - within the phenomenologically
relevant parameter range - very sensitive to the length scale $1/k_j$ of the fluctuations. 
From inspection of  Eq.  (\ref{eq:vorticitywave}), it is also evident that there is a similar sensitivity to
the precise choice of the viscosity. 

In addition to the evolution equations for vorticity, there are equations  for $\theta_T$, $\theta_y$ 
and $d$ that we discuss now.  These describe sound waves and are best solved in Fourier space. 
We concentrate first on a wave traveling in the transverse direction $x_1$, corresponding to $k_1\neq0$, $k_2=k_y=0$. 
In this case, eq.\eqref{eq:hatvartheta_y} decouples from the others
\begin{equation}
\partial_\tau \theta_y + \frac{5}{3\tau} \theta_y + \nu k_1^2 \theta_y =0
\end{equation}
and can be integrated immediately,
\begin{equation}
\begin{split}
\theta_y(\tau,k_1,0,0) = \theta_y(\tau_0,k_1,0,0)
\left(\frac{\tau_0}{\tau}\right)^{5/3} e^{-\nu_0 k_1^2(t-t_0)}.
\end{split}
\end{equation}
Equations \eqref{eq:hatvartheta_T} and \eqref{eq:hatd} are coupled,
\begin{equation}
\begin{split}
\partial_\tau \theta_T - \tfrac{1}{3\tau} \theta_T - k_1^2 d + \nu k_1^2 \left( \tfrac{4}{3} \theta_T + \tfrac{1}{3} 
\theta_y\right)=0,\\
\partial_\tau d + \tfrac{1}{3}(\theta_T + \theta_y) = 0,
\end{split}
\label{eq:41}
\end{equation}
and depend also on the solution for $\theta_y$. Concentrating for simplicity on the case $\theta_y=0$ and eliminating $d$, one finds the second order differential equation
\begin{equation}
\begin{split}
\partial_\tau^2 \theta_T + \left( -\frac{1}{3\tau} + \frac{4}{3} \nu k_1^2 \right) \partial_\tau \theta_T
+\left( \frac{1}{3 \tau^2} 
+ \frac{1}{3} k_1^2 \right) \theta_T =0.
\end{split}
\label{eq:secondorderk1}
\end{equation}
(We have dropped a term suppressed due to $\nu/\tau \ll 1$ in the second bracket.) 
For vanishing viscosity, this equation can be solved in terms of Bessel functions. 
The two linear independent solutions are
\begin{equation}
\tau^{2/3} J_{1/3}\left(\frac{k_1 \tau}{\sqrt{3}}\right) \quad \text{and} \quad \tau^{2/3} Y_{1/3} \left( \frac{k_1 \tau}{\sqrt{3}} \right)\, .
\label{eq:twosolsound_T}
\end{equation}
One finds an oscillating behavior with the amplitude increasing algebraically with time proportional to $(\tau/\tau_0)^{1/6}$. For non-vanishing viscosity $\nu$ there is also an exponential decay which is larger for large wavevectors $k_1$. 
For small values of $k_1 \tau \ll \sqrt{3}$ the two independent solutions of \eqref{eq:secondorderk1} are proportional to $\tau$ and $\tau^{1/3}$, respectively. Eq.\ \eqref{eq:41} implies that the temperature field $d$ grows according to $\tau^2$ and $\tau^{4/3}$ for these two solutions. For $k_1\neq 0$ and late enough times $\tau$, the solutions in \eqref{eq:twosolsound_T} always have an oscillating behavior, however.
For large wave vector $k_1$ one can neglect the terms $\sim \tfrac{1}{\tau}$ and $\sim \tfrac{1}{\tau^2}$ in \eqref{eq:secondorderk1}. Up to viscous damping, the solution corresponds to a perturbation that propagates 
with the velocity of sound $c_S=\sqrt{1/3}$ into the transverse direction.

%
\begin{figure}[t]
\begin{center}
\includegraphics[height=12.0cm, angle=0]{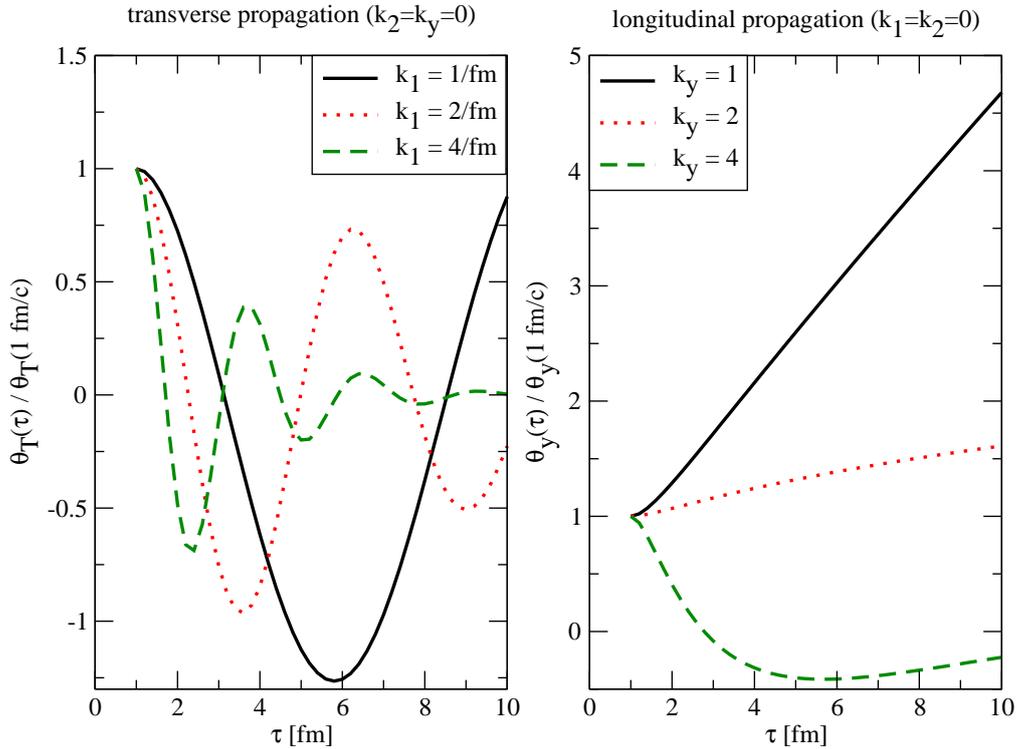}
\caption{
(Left hand side) Velocity divergence amplitude $\theta_T(\tau)/\theta_T(1\text{fm/c})$ for sound 
waves of various wave vectors $k_1$, traveling in the transverse direction $x_1$. The $\tau$-dependence
is calculated from (\ref{eq:secondorderk1}) for $T(1\text{fm/c}) = 500$ MeV and $\eta/s = 1/4\pi$.
(Right hand side) 
 Velocity divergence amplitude $\theta_y(\tau)/\theta_y(1\text{fm/c})$ for a sound wave traveling in the rapidity direction $y$, calculated from  (\ref{eq:47}) for the same input values.
}
\label{fig68}
\end{center}
\end{figure}

We show the time evolution governed by (\ref{eq:secondorderk1}) on the left hand side of  Fig.\ \ref{fig68}. 
In close similarity to the case of vorticity, we observe that also fluctuations in the transverse velocity
divergence can persist over time scales relevant in heavy ion collisions. For sound waves in the transverse
direction, the fluid acts like an efficient low-pass filter, allowing for the unattenuated (or even slightly
amplified) passage of fluctuations on sufficiently large scales $1/k_1 \geq 1$ fm, while filtering out
fluctuations on smaller length scales $1/k_1 < 0.25$ fm. 

We finally turn to sound waves traveling in the rapidity direction $y$, i.\ e.\ $k_y\neq 0$, $k_1=k_2=0$. Now, equation \eqref{eq:hatvartheta_T} decouples 
\begin{equation}
\partial_\tau \theta_T - \tfrac{1}{3\tau} \theta_T + \nu \tfrac{1}{\tau^2} k_y^2 \theta_T = 0
\end{equation}
and can be integrated immediately,
\begin{equation}
\theta_T(\tau,0,0,k_y) =  \theta_T(\tau_0,0,0,k_y) 
 \left(\frac{\tau}{\tau_0}\right)^{1/3} e^{\nu_0 \frac{9}{8\sqrt{t_0}}k_y^2\left(\tfrac{1}{\sqrt{t}}-\tfrac{1}{\sqrt{t_0}}\right)}\, .
\end{equation}
The equations for $\theta_y$ and $d$ are coupled and depend on $\theta_T$,
\begin{equation}
\begin{split}
\partial_\tau \theta_y + \tfrac{5}{3\tau} \theta_y - \tfrac{1}{\tau^2} k_y^2 d+\nu \tfrac{1}{\tau^2}k_y^2 \left( \tfrac{4}{3} \theta_y + \tfrac{1}{3} \theta_T \right) &=0\, ,\\
\partial_\tau d + \tfrac{1}{3} (\theta_T + \theta_y) &=0\, .
\end{split}
\label{eq:46}
\end{equation}
Concentrating on $\theta_T=0$, this yields the following second order differential equation for $\theta_y$ (we drop again a term suppressed due to $\nu/\tau \ll 1$),
\begin{equation}
\partial_\tau^2 \theta_y + \left(\frac{5}{3\tau} + \nu \frac{4}{3\tau^2} k_y^2\right) \partial_\tau \theta_y + \left(-\frac{5}{3 \tau^2}+\frac{1}{3 \tau^2} k_y^2\right) \theta_y=0.
\label{eq:47}
\end{equation}
The two linear independent solutions to this equation for $\nu=0$ are
\begin{equation}
\tau^{-\tfrac{1}{3}+\tfrac{1}{3}\sqrt{16-3 k_y^2}} \quad \text{and} \quad \tau^{-\tfrac{1}{3}-\tfrac{1}{3}\sqrt{16-3 k_y^2}}.
\label{eq:soundrapidity}
\end{equation}
For $k_y^2\ll 16/3$ this becomes $\tau$ and $\tau^{-5/3}$. Equation \eqref{eq:46} implies that this corresponds to perturbations in the temperature field $d$ that grow like $\tau^2$ or decay like $\tau^{-2/3}$, respectively.
For $k_y^2>16/3$ the solutions \eqref{eq:47} 
correspond to an oscillation with a period that increases as a function of time and an additional decrease in the amplitude. 
On the right hand side of Fig.~\ref{fig68}, we  plot this behaviour of $\theta_y$ for small non-vanishing
viscosity, when the algebraic increase is modified by an exponential decay. We observe again that 
fluctuations of modes with wave vectors $k_y \sim O(1)$ can persist 
or can be amplified over time scales commensurate with the expected expansion duration of heavy ion collisions. 

In the limit of large $k_y^2$, one can translate \eqref{eq:soundrapidity} back to position space and one finds that it corresponds to an excitation that propagates in the rapidity direction according to
\begin{equation}
\frac{\partial y}{\partial \tau} = \frac{c_S}{\tau} = \frac{1}{\tau} \sqrt{\frac{1}{3}}.
\end{equation}
In general, for a sound mode propagating fastly into an arbitrary direction described by a large wavevector $k=(k_1,k_2,k_y)$, we expect the propagation velocity $(\frac{\partial x_1}{\partial\tau},\frac{\partial x_2}{\partial\tau},\frac{\partial y}{\partial\tau})$ to satisfy
\begin{equation}
\left(\frac{\partial x_1}{\partial\tau}\right)^2+
\left(\frac{\partial x_2}{\partial\tau}\right)^2+
\tau^2 \left(\frac{\partial y}{\partial\tau}\right)^2
= \frac{1}{3}.
\label{sound}
\end{equation}
This condition is satisfied by the sound modes in $\theta_y$ and $\theta_T$ discussed here.

\section{Turbulent fluctuations}
\label{sec5}

In full generality, the non-linear equations \eqref{eq:A} and \eqref{eq:C} or, equivalently, \eqref{eq:velocityeqtransformed} and \eqref{eq:deqtransformed} are difficult to analyze. One can always use a splitting of the velocity into a solenoidal part that carries vorticity $\omega_j$ and an irrotational part described by the divergence $\theta$.  The nonlinear terms in the equation of motion will lead to couplings between these fields and to the logarithmic temperature field $d$, however. Intuitively, one expects that perturbations in the fluid divergence propagate quickly in the medium with the characteristic velocity given by the velocity of sound. 
The fluid velocity $(u^1,u^2,u^y)$ can be small compared to sound propagation.  
This is often characterized by a small Mach number
\begin{equation}
\text{Ma} = \frac{\sqrt{(u^1)^2+(u^2)^2+\tau^2(u^y)^2}}{c_S} 
\ll 1\, .
\label{eq:Machnumber}
\end{equation}
For the description of the part of the fluid velocity that carries vorticity one can often assume in this case a vanishing divergence, $\theta=0$, since the fast sound modes can be viewed as decoupled from the slower modes dominating the solenoidal part of the fluid velocity.
In the present section we will study the equations of motion in this situation and show that there are some interesting parallels to non-relativistic, incompressible fluids.

For our discussion in this section, it will be useful to work with the rescaled velocities $v_i$ introduced in 
section~\ref{sec3}. For $\theta=\vartheta=0$, Eq.\ \eqref{eq:velocityeqtransformed} simplifies then to
\begin{equation}
\partial_t v_j + \sum_{m=1}^2 v_m \partial_m v_j + \frac{1}{\tau^2} v_y\partial_y v_j +\partial_j d
-\nu_0 \left( \partial_1^2 + \partial_2^2 + \frac{1}{\tau^2} \partial_y^2 \right) v_j =0.
\label{eq:velocityeqv}
\end{equation}
Due to the the solenoidal constraint
\begin{equation}
\partial_1 v_1 + \partial_2 v_2 + \frac{1}{\tau^2} \partial_y v_y =0
\label{eq:solenoidalconstraintv}
\end{equation}
the temperature field $d$ is not independent of the velocity field. More specific, by taking the divergence of \eqref{eq:velocityeqv} one derives
\begin{equation}
\left( \partial_1^2 + \partial_2^2 + \frac{1}{\tau^2} \partial_y^2 \right) d + \sum_{m,n=1}^2 (\partial_m v_n)(\partial_n v_m) 
+ \frac{2}{\tau^2} \sum_{m=1}^2 (\partial_m v_y)(\partial_y v_m) + \frac{1}{\tau^4} (\partial_y v_y)^2 = 0.
\label{eq:connectiondtov}
\end{equation}
This is an instant of the Poisson equation for $d$. For given boundary conditions it can be inverted to yield $d$ as a (non-local) functional of the velocity field.

Equation \eqref{eq:velocityeqv} has some interesting features~\footnote{
We mention as an aside that in addition to the obvious rotation symmetry in transverse direction and the translational symmetries in transverse and rapidity directions, Eqs.\ \eqref{eq:velocityeqv} and \eqref{eq:solenoidalconstraintv} have also the following space-time symmetry
\begin{equation}
\begin{split}
x_m & \to x_m + V_m t \quad (m=1,2),\\
y & \to y - 2 V_y \frac{t}{\tau^2},\\
v_j & \to v_j + V_j \quad (j=1,2,y)
\end{split}
\end{equation}
for constant velocity $V_j$. Indeed, contributions from the time derivative and from the non-linear advection terms cancel. For $V_y=0$ this corresponds to Galilean symmetry in the transverse plane while the situation is more complicated for $V_y\neq0$. However, it is not so clear whether this invariance is very useful since there is no translational symmetry with respect to time.}.
It takes the form of a two-dimensional Navier-Stokes equation in situations where $v_y=0$ and where the dependence of $v_1, v_2$ on $y$ can be neglected. Moreover, for a large class of initial conditions at time $\tau=\tau_0$, the evolution becomes effectively two-dimensional for late times $\tau/\tau_0 \gg 1$. Indeed, both the non-linear velocity term that couples $v_y$ to $v_1$ and $v_2$ and the damping term involving the derivative with respect to rapidity contain factors that decrease as $1/\tau^2$.
Similarly, the solenoidal constraint \eqref{eq:solenoidalconstraintv} assumes its two-dimensional form in that limit.

Motivated by the observation that for the particular expansion geometry of the Bjorken model,
fluctuations are governed by an evolution equation that reduces at late times to a
non-relativistic Navier-Stokes equation (in rescaled time coordinates),
we now discuss the conditions for a non-linear turbulent evolution of fluctuations by the very
concepts that have proven useful in the classification of solutions of the Navier-Stokes equation. 
To this end, we consider situations where the velocities change notably on distances of order $l$
 in the transverse direction or for rapidity differences $\Delta y$. The damping term in Eq.\ \eqref{eq:velocityeqv} leads then to a damping rate (inverse relaxation time) that can be characterized in terms of
 the dimensionless number 
 \begin{equation}
\kappa=\frac{l^2}{\tau^2 \Delta y^2}\, .
\label{eq:definitionkappa}
\end{equation}
This damping rate is of order $\nu_0/l^2$ for $\kappa\ll 1$, and it is of order
 $\nu_0/(\tau^2 \Delta y^2)$ for $\kappa\gg1$.
For characteristic velocities $v_T$ in transverse, respectively $v_y$ in rapidity direction, the flow can then be characterized in terms of the Reynolds numbers 
\begin{equation}
\text{Re}^{(T)} = \frac{v_T l}{\nu_0}, \quad \text{Re}^{(y)} = \frac{v_y\; l^2}{\nu_0\;\Delta y} \frac{1}{\tau^2}
\qquad \text{for}\quad \kappa \ll 1
\label{eq:Reynolds1}
\end{equation}
and
\begin{equation}
\text{Re}^{(T)} = \frac{v_T \;\tau^2 \;\Delta y^2}{\nu_0 \;l}, \quad \text{Re}^{(y)} = \frac{v_y\; \Delta y }{\nu_0}
\qquad \text{for}\quad \kappa \gg 1\, .
\label{eq:Reynolds2}
\end{equation}
Obviously, these definitions can be extended to intermediate $\kappa$, as well.

If both Reynolds numbers are small, $\text{Re}^{(T)}\ll 1$, $\text{Re}^{(y)}\ll 1$, the resulting flow pattern is expected to be laminar. The viscous damping term dominates then over the nonlinear terms in Eq.\ \eqref{eq:velocityeqv} and velocities are expected to follow a regular behavior dominated by an exponential decay in time. More specifically, if the non-linear term in Eq.\ \eqref{eq:velocityeqv} can be neglected one can use Fourier decomposition with respect to the spatial arguments and one finds the solution
\begin{equation}
v_j(t,k_1,k_2,k_y)  =  v_j(t_0,k_1,k_2,k_3)   e^{-\nu_0(k_1^2+k_2^2)(t-t_0)+\nu_0 \tfrac{9}{8 \sqrt{t_0}}k_y^2 \left(\tfrac{1}{\sqrt{t}}-\tfrac{1}{\sqrt{t_0}}\right)}.
\end{equation}
This resembles closely the behavior of vorticity in Eq.\ \eqref{eq:vorticitywave}.

In contrast, if both Reynolds numbers are large, $\text{Re}^{(T)} \gg 1$, $\text{Re}^{(y)} \gg 1$, one expects a turbulent regime. The nonlinear terms dominate now over the viscous damping term in Eq.\ \eqref{eq:velocityeqv} and the velocities will change rather irregularly from point to point both in the transverse plane and for different values of the rapidity variable $y$. 

We consider next the case $\text{Re}^{(T)}\ll 1$ and $\text{Re}^{(y)} \gg 1$. For $\kappa\gg 1$ the derivatives with respect to $x_1$ and $x_2$ effectively drop out from Eq.\ \eqref{eq:velocityeqv}. One might expect a sort of one-dimensional turbulent behavior. However, the unusual explicit time dependence of the non-linear and damping terms could spoil this conclusion. Also for $\kappa \ll 1$, we are unable to predict consequences of \eqref{eq:velocityeqv} without additional explicit calculations. However, because of the late time behavior
of $\text{Re}^{(y)} \gg 1$, we do not expect that this case is particularly relevant for the simulation of
heavy ion collisions (see the more detailed argument in the paragraph below).

In the opposite case $\text{Re}^{(T)}\gg 1$, $\text{Re}^{(y)}\ll 1$, and for $\kappa\ll 1$, all terms containing derivatives with respect to $y$ can be neglected and Eq.\ \eqref{eq:velocityeqv} can be seen as a set of equations describing fluid motion in $2+1$ dimensions with rapidity entering only as a parameter.  With respect to the transverse coordinates one would expect a turbulent flow pattern.
We remark that $\kappa$ as defined in Eq.\ \eqref{eq:definitionkappa} decreases with time $\tau$ and
 the ratio of the two Reynolds numbers (which is independent of $\kappa$)  increases with time
\begin{equation}
\text{Re}^{(T)} / \text{Re}^{(y)} = \frac{v_T \, \Delta y \;\tau^2}{v_y\; l }\, .
\label{eq5.10}
\end{equation}
Therefore, in contrast to the case $\text{Re}^{(T)}\ll 1$, $\text{Re}^{(y)} \gg 1$ discussed above,
the case $\text{Re}^{(T)}\gg 1$, $\text{Re}^{(y)}\ll 1$ can persist at late times, and it can be reached dynamically.
For completeness, we mention finally the region $\kappa\gg 1$ when derivatives with respect to $y$ enter in the damping term. Considered as a function of the transverse coordinates $x_1$ and $x_2$, the velocity field will be damped by locally varying rates due to irregularities with respect to the rapidity argument. However, this damping is dominated by the first non-linear term in Eq.\ \eqref{eq:velocityeqv} so that the fluid will again behave turbulent with respect to the transverse coordinates.

The above discussion indicates that the case $\text{Re}^{(T)}\gg 1$, $\text{Re}^{(y)}\ll 1$, $\kappa\ll 1$ may be of particular relevance for the discussion of the onset of non-linear turbulent behavior in heavy ion collisions, since it
can be reached dynamically and since it persists at late times. {\it For a large set of initial conditions, we therefore
expect that the fluid dynamics of heavy ion collisions evolves towards a system with effectively two-dimensional
turbulent behavior.} To address the issue to what extent the evolution towards turbulence  could be completed
within the finite duration of a heavy ion collision, let 
us finally put some numbers to this parametric discussion. Because of the late time limit of eq.~(\ref{eq5.10}),
we focus on the transverse Reynolds number
\begin{equation}
\text{Re}^{(T)} = \frac{v_T\, l\, T\, s}{\eta}.
\end{equation}
We consider  typical values for a heavy ion collision at LHC energies, e.g. 
$T \approx 0.3\, \text{GeV}$ and a  length scale $l \approx 5 \, \text{fm}$. Taking for the transverse 
velocity a fraction of the velocity of light, say $v_T=0.1 \, c$, one finds 
$v_T \, l\, T \approx 1$ and
\begin{equation}
\text{Re}^{(T)}=\frac{1}{\eta/s}\, .
\end{equation}
For small values of the normalized viscosity $\eta/s<1$, one therefore expects a transverse Reynolds number 
$\text{Re}^{(T)}$  that is larger than unity but not many orders of magnitude larger than unity
($\text{Re}^{(T)} < 100$). Such values are not sufficiently large to expect
fully developed turbulence. A value $\text{Re}^{(T)} > 1$ indicates,
however, that the system can be driven outside the
region of validity of a laminar evolution and that it may display features indicative of the onset of turbulent
behavior.

\section{Qualitative features of turbulence}
\label{sec6}
In the previous section, we have discussed the general conditions under which the time-evolution
of fluctuations in a Bjorken expansion scenario can be expected to lead to the onset of turbulent
behavior. Despite the relativistic nature of the system under study, we found that upon coordinate
transformation, the time evolution of fluctuations is governed by an equation that takes the form of
a two-dimensional Navier-Stokes equation at late times. Although the Navier-Stokes equation
presents still many deep problems, much is known about fully developed turbulence in this system. 
Classical achievements include in particular the scaling theory by Kolmogorov \cite{Kolmogorov} for three-dimensional turbulence  and its extension to the two-dimensional case mainly by Kraichnan \cite{Kraichnan} 
and Batchelor \cite{Batchelor}. For reviews of this field, see~\cite{TurbulenceReviews}

To the best of our knowledge, the Bjorken scenario considered here is the first example of a relativistically 
expanding three-dimensional scenario that under suitable initial conditions evolves dynamically into a
system with effectively two-dimensional turbulent dynamics described by a non-relativistic 
Navier-Stokes equation. Turbulence in the two-dimensional case is known to display some characteristic
qualitative differences in comparison to the three-dimensional case. Although the present section contains,
strictly speaking, no novel results, our finding of an  effectively
two-dimensional fluid dynamic propagation of fluctuations at late times prompts us 
to discuss here the pertinent features of fully
developed turbulence that apply to the Bjorken scenario for sufficiently large values of $\text{Re}^{(T)}$. 
In particular, we point to the phenomenon of an inverse cascade that exists only in the case of
two-dimensional turbulence and that provides a unique mechanism for enhancing fluctuations on large 
spatial scales during the evolution towards turbulence. 
To set the stage for this discussion, we introduce first shortly a statistical description of fluctuations in the fluid velocities and
energy densities. 

\subsection{Fluctuation spectra}
\label{sec6.1}
In general, fluctuations in the fluid velocities and the energy densities
can be described statistically in terms of a $\tau$-dependent probability distribution
\begin{equation}
p_{\tau}[u^\mu(\tau,x_1,x_2,y), \epsilon(\tau,x_1,x_2,y)]\, .
\label{eq:probdist}
\end{equation}
Eq.\ \eqref{eq:probdist} describes an ensemble of events with equal ``macroscopic'' properties such as nucleon number, center of mass energy and impact parameter. Bjorkens model of a unique fluid velocity and energy density  is then recovered by taking the probability distribution in \eqref{eq:probdist} to be infinitely narrow.

We consider a generalization of Bjorkens model where the assumed symmetries (boost invariance in the longitudinal direction and translation and rotation invariance in the transverse plane) are broken by the fluctuations for a particular event but hold in a {\itshape statistical} sense. This means that the probability distribution \eqref{eq:probdist} is invariant under these symmetries. This implies in particular that the expectation value of velocity is given by
\begin{equation}
\langle u^\mu \rangle = (1,0,0,0)
\end{equation}
for all times $\tau>\tau_0$ and for all values of $x_1,x_2,y$. Similarly the expectation values 
of thermodynamic scalars such as $\epsilon$, $p$, $s$, $T$ etc. will be a function of $\tau$, only. In general, however, the $\tau$-dependence of these expectation values will differ from the ones obtained by Bjorken due to non-linear effects of fluctuations.

Beyond the expectation values, the probability distribution \eqref{eq:probdist} can be characterized in terms of correlation functions. In particular, the two-point correlation function of velocities at equal time $\tau$ is defined by ($i,j=1,2,y$)
\begin{equation}
G_u^{ij} (\tau, x_1-x_1^\prime, x_2-x_2^\prime, y-y^\prime)
= \langle u^i(\tau,x_1,x_2,y) \; u^j(\tau,x_1^\prime,x_2^\prime,y^\prime) \rangle.
\label{eq:Guij}
\end{equation}
Due to translational symmetries this depends only on the differences in the spatial coordinates. Similarly, we define
\begin{equation}
G_T (\tau, x_1-x_1^\prime, x_2-x_2^\prime, y-y^\prime)
= \langle T(\tau,x_1,x_2,y) \; T(\tau,x_1^\prime,x_2^\prime,y^\prime) \rangle
\label{eq:Gepsilon}
\end{equation}
and the cross-correlation
\begin{equation}
G_{uT}^j (\tau, x_1-x_1^\prime, x_2-x_2^\prime, y-y^\prime)\\
= \langle u^j(\tau,x_1,x_2,y) \; T(\tau,x_1^\prime,x_2^\prime,y^\prime) \rangle.
\label{eq:Guepsilonj}
\end{equation}
The generalization to other scalar quantities such as energy density $\epsilon$ or pressure $p$ or to un-equal time arguments is obvious. It is useful to introduce also the Fourier decomposition
\begin{equation}
G_u^{ij}(\tau,x_1,x_2,y) 
= \int \frac{d^3 k}{(2\pi)^3} e^{i(k_1 x_1+k_2 x_2+k_y y)} \tilde G_u^{ij}(\tau,k_1,k_2,k_y)
\end{equation}
and the abbreviation
\begin{equation}
G_u^{ij}(\tau)=G_u^{ij}(\tau,x_1=0,x_2=0,y=0)\, .
\end{equation}
This generalizes trivially to the other functions. We note that for symmetry reasons, one has $G_u^{ij}(\tau)=0$ for $i\neq j$ and $G_u^{11}(\tau)=G^{22}(\tau)$.

For a solenoidal fluid with $\partial_j u^j=0$ one has
\begin{equation}
\partial_i G^{ij}_u(\tau,x_1,x_2,y)=0
\end{equation}
or in momentum space
\begin{equation}
k_i G^{ij}_u(\tau,k) =0.
\end{equation}

In this framework, characterizing the fluid evolution of a heavy ion collision amounts to make statements about
the form of correlation functions such as  $G_u^{ij}(\tau,k)$ etc.\ at some given time $\tau$.   It is clear that the form of these correlations depends strongly on the initial conditions. For example, if the initial fluctuations at time $\tau_0$ are small enough (or, equivalently, the viscosities large enough to have small Reynolds numbers) so that the linearized equations \eqref{eq:hatvartheta_T} - \eqref{eq:hatomega} can be applied, one can derive from them linear evolution equations for the set of correlation functions $G_u^{ij}$, $G_d$ and $G_{ud}^j$. The form of these functions at time $\tau$ is then directly linked to the corresponding functions at time $\tau_0$.

The situation is much more complicated in the presence of non-linear contributions to time evolution, even if the system is far from the conditions of fully developed turbulence. Indeed, if one tries to use the non-linear equations \eqref{eq:A} and \eqref{eq:C} to derive evolution equations for the two-point correlation functions, one finds that three-point correlations get involved, as well. The evolution equation for these involve even higher correlations and so on. This is an instant of the well-known closure problem in the statistical description of fluids.  The same problem appears in non-perturbative formulations of quantum and statistical field theories.
 No exact analytical solutions are known and advanced techniques from statistical mechanics and field theory are needed to find approximate ones. The scaling theory of
Kolmogorov provides some insight into these problems.


\subsection{Scaling theory of turbulence at large Reynolds number}
Any fluctuation present in the initial conditions of a relativistic heavy ion collision can be regarded
as a source of non-thermal, say 'mechanical' energy. Most generally, one would like to understand to
what extent this mechanical energy dissipates to thermal energy, and to what extent it does not 
but leaves characteristic, dynamically evolved fluctuations  visible in final state observables. 
In section~\ref{sec4}, we have shown examples of the dissipation
(or amplification) of some fluctuation modes in a linearized description that applies to very small
Reynolds numbers. Here, we want to comment on the opposite case of very large Reynolds number or very
small viscosity. For large Reynolds numbers, we have found in section~\ref{sec5} that fluctuations on 
a Bjorken background field are governed by an evolution equation that takes for sufficiently late
times the form of a non-relativistic  Navier-Stokes equation. Also,  for situations of sufficiently small
Mach number, there is a rationale for setting $\theta \to 0$. Therefore, our discussion of the case
of large Reynolds number will reduce to recalling pertinent features of  
Kolomogorov's theory of homogeneous turbulence for a non-relativistic, incompressible fluid 
at very large Reynolds numbers.

For a non-relativistic fluid, it is common to denote the fluid kinetic energy per unit mass 
by $\tfrac{1}{2}[\vec v^2]$, where the square brackets stand for spatial averaging. 
The rate of dissipation to thermal energy for a three-dimensional evolution is given by
\begin{equation}
\varepsilon_{\rm diss} = \frac{d}{dt} \frac{1}{2}[ \vec v^2 ] = \frac{1}{2} \nu [ \sum_{i,j=1}^3 (\partial_i v_j+\partial_j v_i)^2 ]= \nu [ \vec \omega^2 ]\, .\qquad \text{(3-dim. case)}
\end{equation}
This shows that energy dissipation is mainly due to fine structures of the velocity field for which the gradients are large; dissipation is mainly taking place at large wave-vectors $k$. If one now decreases the viscosity, one finds finer and finer structures emerge so that the energy dissipation rate $\varepsilon_{\rm diss}$ remains positive. In fact, the
mean-square vorticity $\tfrac{1}{2}[ \vec \omega^2 ]$ (also called enstrophy) grows $\sim 1/\nu$ for $\nu\to 0$. This is possible since $\tfrac{1}{2}[ \vec \omega^2 ]$ is not only given by the vorticity present at some initial time or generated from an external driving force. Rather, it can be generated also by non-linear terms in the three-dimensional Navier-Stokes equation. The mechanical energy is cascaded from the large length structures to the smaller ones by virtue of the non-linear terms in the Navier-Stokes equation. This is the famous cascade picture of Richardson~\cite{Richardson}. In his words: ``Big whorls have little whorls, Which feed on their velocity; And little whorls have lesser whorls, And so on to viscosity.''

Let us now come to the situation in two spatial dimensions. The most important difference to the three-dimensional case concerns the evolution of mean-square vorticity in the absence of an external driving force (vorticity $\omega = \partial_1 v_2 - \partial_2 v_1$ has now only a single component)
\begin{equation}
\frac{d}{dt} \frac{1}{2} [ \omega^2 ] = -\nu[ (\vec \nabla \omega)^2 ]\, . \qquad \text{(2-dim. case)}
\end{equation}
This shows that $\tfrac{1}{2} [ \omega^2 ]$  never increases as a function of time. This in turn implies that energy per unit mass is conserved for vanishing vorticity,
\begin{equation}
\frac{d}{dt} \frac{1}{2} [ \vec v^2 ] \to 0 \quad\quad \text{for} \quad\quad \nu\to 0.
\end{equation}
A cascade of mechanical energy from large structures into smaller ones where it is finally dissipated is therefore not possible \footnote{It has been argued that a cascade can take place for enstrophy $\tfrac{1}{2}[ \omega^2 ]$ instead of energy $\tfrac{1}{2}[ \vec v^2 ]$, however. This is possible if $[ (\vec \nabla\omega)^2 ]$ grows $\sim 1/\nu$ for $\nu\to 0$ due to non-linear terms.}.

Without going further into the detailed mechanism of two-dimensional turbulence let us now attempt to transfer some of the insights that have been gained in this field to heavy ion physics, in particular fluctuations around Bjorken flow. We consider the case of large fluctuations and small kinematic viscosity $\nu$ so that the Reynolds number $\text{Re}$ is large. Also, we concentrate on the limit of large time $\tau$ where \eqref{eq:velocityeqv} becomes two-dimensional. Similar to the non-relativistic case one has now
\begin{equation}
\frac{d}{dt} [ v_1^2+v_2^2 ] \to 0 \quad \quad \text{for} \quad \quad \nu \to 0\, .
\end{equation}
For a fixed time $t$ and rapidity $y$ one can characterize the correlations of the velocities in the transverse plane by ($m,n = 1,2$)
\begin{equation}
\begin{split}
(G_v)_{mn} (t,x_1-x_1^\prime, x_2-x_2^\prime,0) 
& = \langle v_m(t,x_1,x_2,y) v_m(t,x_1^\prime,x_2^\prime,y) \rangle \\
& = \left(\frac{\tau_0}{\tau} \right)^{2/3} G_u^{mn}(\tau,t,x_1-x_1^\prime, x_2-x_2^\prime,0)\, .
\end{split}
\label{eq:78}
\end{equation}
Adapting to the standard notation used in the literature about turbulence, we write the Fourier transform of this as
\begin{equation}
(G_v)_{mn}(t,x_1,x_2,y) = \int \frac{d^2 k}{(2\pi)^2} e^{i(k_1 x_1+k_2 x_2)}
 \left(\delta_{mn}-\frac{k_m k_n}{k_1^2+k_2^2} \right) \frac{2\pi}{k} E(t,k).
\label{eq:tvsft}
\end{equation}
The tensor structure of \eqref{eq:tvsft} follows from rotational invariance and from the solenoidal constraint \eqref{eq:solenoidalconstraintv}. The function $E(t,k)$ depends on $k_1$ and $k_2$ only in the combination $k=\sqrt{k_1^2+k_2^2}$. The normalization in \eqref{eq:tvsft} is chosen such that
\begin{equation}
E(t) = \frac{1}{2} \langle v_1^2 + v_2^2 \rangle = \frac{1}{2} \sum_{m=1}^2 (G_v)_{mm}(t,0,0,0)
 = \int_0^\infty d k\;  E(t,k).
\label{eq:zuz2}
\end{equation}
For a non-relativistic fluid, the function $E(t,k)$ describes how the fluid kinetic energy per unit mass is distributed over the different wave vectors. We emphasize that in the relativistic setup considered here, $E(t,k)$ is not directly representing kinetic energy. Instead it simply parameterizes the contribution to the fluctuating transverse velocity field from different wave vectors. The contribution of these fluctuations to kinetic energy, for example in the laboratory frame, can be determined but the resulting relation is more-complicated than in the non-relativistic case. 

For the case of a relativistic heavy ion collision, the initial distribution $E(t_0,k)$ would characterize the 
relative strength with which different length scales $1/k$ are represented in the fluctuating initial conditions. 
The question of how this distribution of fluctuations evolves amounts then to studying the time-dependence
of $E(t,k)$. Here, we point only to one remarkable feature of the time-dependence of a two-dimensional
fluid at large Reynolds number, that can be understood in the scaling theory of freely decaying turbulence in two dimensions developed by Batchelor \cite{Batchelor}. This theory is based on the assumption that at
sufficiently late times, the function $E(t,k)$ remembers only a single number from its initialization, namely
its average fluid velocity $\lambda$ defined by $\lambda^2 = \frac{1}{2} \langle v_1^2+v_2^2 \rangle$.
From dimensional reasoning it follows then that
\begin{equation}
E(t,k) = \lambda^3 t \;h(k\lambda t)\, .
\end{equation}
It follows from \eqref{eq:zuz2} that the dimensionless function $h(x)$is normalized to unity, 
$\int_0^\infty dx \; h(x) = 1$. Interestingly, if one assumes that $E(t,k)$ is dominated by the region 
around some characteristic wave vector $k_M$, then this scale will change with time according to
\begin{equation}
k_M \sim \frac{1}{\lambda t}.
\end{equation}
{\it This implies that kinetic energy is shifted from small length scales to larger ones}, in contrast to the Richardson cascade in three dimensions~\cite{Richardson}. We would like to close this section on a cautious but speculative note:
We recall first that Batchelor's theory of freely decaying turbulence was
developed for very large Reynolds numbers that may not be realized in 
heavy ion collisions. However, the above considerations
may make it conceivable that non-linear effects in the fluid
dynamic evolution related to the onset of turbulence can move fluctuations in the initial kinetic energy
to larger spatial scales. This phenomenon would be a distinct characteristics of an effectively two-dimensional
turbulent evolution, and it may be identified experimentally by finding fluctuations related to length scales
that are inconceivable to be present in fluctuating initial conditions.

\section{The sensitivity of particle spectra on velocity correlation functions}
\label{sec7}
In the previous section~\ref{sec6}, we have seen that the fluid dynamic evolution of fluctuations
can be characterized efficiently in terms of correlation functions of fluid velocities. Here we discuss
how information about such velocity correlations enters the particle spectra that are experimentally
accessible in heavy ion collisions. 

The starting point of our discussion is the 'freeze-out' phase space 
distribution $f(x,p)$ that parametrizes the matter distribution at the time 
$\tau_{\rm fo}$ when the particles decouple from the fluid dynamic evolution.  
In the following, we shall view $f(x,p)$ as describing an event-averaged smooth fluid
system supplemented by event-specific fluctuations. We shall then ask how these fluctuations
are reflected in observables.  This logic should apply to arbitrary choices of $f(x,p)$. For 
the purpose of illustration, however, we shall restrict our discussion to a simple ansatz~\footnote{We 
note as an aside that phenomenologically more realistic choices
of $f(x,p)$ would be significantly more complex. In particular, they would contain information about
the finite spatial extent of the matter distribution in the transverse and longitudinal distribution, 
they would supplement the ideal gas expression  \eqref{eq:idealBoltzm} by terms proportional to 
viscosity~\cite{Heinz:2009xj,Romatschke:2009im,Teaney:2009qa}, 
and they may implement correct Fermi-Dirac or Bose-Einstein statistics 
instead of the Boltzmann distribution \eqref{eq:idealBoltzm}. } 
that describes a locally approximately thermal
distribution consistent with the Bjorken background field of section~\ref{sec3}, and that allows for
the implementation of local fluctuations on top of this background field. A simple choice with
these properties is the Boltzmann distribution
\begin{equation}
f(x,p) = d \;e^{\tfrac{p_\mu u^\mu(x)}{T(x)}}\, ,
\label{eq:idealBoltzm}
\end{equation}
where the normalization $d$ is fixed by the spin and flavor degeneracy of the degrees of freedom
that decouple from the system. Assuming that the freeze-out takes place at some proper time 
$\tau_\text{fo}$ when the average temperature drops below some freeze-out temperature 
$T_\text{fo}$, the hadronic spectra can be calculated 
using the Cooper-Frye freeze-out prescription
\begin{equation}
E\frac{dN}{d^3 p} = \int \frac{p_\mu d\Sigma^\mu}{(2\pi)^3} f(x,p)\, .
\label{eq:CFfreezeout}
\end{equation}
Here, the freeze-out volume is determined by $p_\mu d\Sigma^\mu = m_T \tau \text{cosh}(\eta-y) dx_1 dx_2 dy$
for the case of Bjorken expansion. 

In practice, the spectra (\ref{eq:CFfreezeout}) measured in heavy ion collisions include averaging 
over many events. On the level of the freeze-out distribution $f(x,p)$, this event average corresponds  
to an ensemble average with respect to the fluid velocity, energy 
density or temperature fields, respectively. Denoting the corresponding averaging 
by triangular brackets, we replace therefore in the calculation of (\ref{eq:CFfreezeout})
the function $f$ by
\begin{equation}
f(x,p) = d {\bigg\langle} e^{\tfrac{p_\mu u^\mu(x)}{T(x)}} {\bigg\rangle}.
\end{equation}
Denoting the particle four-momentum by $(p^0,p^1,p^2,p^3) = (m_T \,\text{cosh}\,y, p^1,p^2,m_T \,\text{sinh}\,y)$ with transverse mass squared $m_T^2=p_T^2+m^2$, $p_T^2=(p^1)^2+(p^2)^2$, we expand 
$f(x,p)$  for small fluctuations around the velocity profile of Bjorken 
$(u^0,u^1,u^2,u^y) =\left(\cosh\eta,0,0,\sinh\eta\right)$ and the constant freeze-out temperature $T_\text{fo}$
\begin{equation}
\begin{split}
f&(x,p) = d\; e^{-\frac{m_T\,\text{cosh}(\eta-y)}{T_\text{fo}}} {\bigg \langle} 1+ \frac{m_T \,\text{cosh}(\eta-y)}{T_\text{fo}^2}(T-T_\text{fo})
 + \frac{1}{T_\text{fo}} \left( p_1 u^1+p_2 u^2+\tau\,m_T\, \text{sinh}(\eta-y) \;u^y \right) \\
& +\frac{1}{2T_\text{fo}^2} \left( p_1 u^1+ p_2 u^2 + \tau\,m_T\,\text{sinh}(\eta-y) \;u^y \right)^2
 + \left( \frac{m_T^2\, \text{cosh}^2 (\eta-y)}{2\,T_\text{fo}^4}-\frac{m_T\,\text{cosh}(\eta-y)}{2\,T_\text{fo}^3} \right)(T-T_\text{fo})^2\\
& + \frac{m_T\,\text{cosh}(\eta-y)}{T_\text{fo}^4}(p_1 u^1 + p_2 u^2 + \tau\,m_T\,\text{sinh}(\eta-y) \;u^y)
 (T-T_\text{fo}) +\dots {\bigg \rangle}.
\label{eq:Bdexpansioninfluct}
\end{split}
\end{equation}
The terms linear in $T-T_\text{fo}$ or $u^j$ vanish by definition or due to symmetry reasons and similar the cross-terms $\sim u^j u^i$ with $i\neq j$. Also, due to translational symmetry the variances at $\tau=\tau_\text{fo}$ are actually independent of the coordinates $x_1, x_2$ and $y$, such that
\begin{equation}
\begin{split}
&\langle (u^1)^2 \rangle  = G_u^{11}(\tau_\text{fo}),\\
&\langle (u^y)^2 \rangle  = G_u^{yy}(\tau_\text{fo}),\\
&\langle (T-T_\text{fo})^2 \rangle = G_T(\tau_\text{fo}).
\label{eq7.5}
\end{split}
\end{equation}
This allows one to write the one-particle spectrum for small fluctuations around Bjorken flow as
\begin{equation}
E \frac{dN}{d^3p} =  E {\bigg [} \frac{d N_0}{d^3 p} + \frac{d \delta N_1}{d^3 p} G_u^{11}(\tau_\text{fo}) + \frac{d\delta N_2}{d^3 p} G_u^{yy}(\tau_\text{fo}) 
 +\frac{d\delta N_3}{d^3p} G_T(\tau_\text{fo}) {\bigg ]},
\end{equation}
with
\begin{equation}
\begin{split}
E \frac{dN_0}{d^3 p} = & \frac{d \, \tau_\text{fo}\, R_0^2 \,m_T}{4\pi^2} \,K_1\left(\frac{m_T}{T}\right),\\
E \frac{dN_1}{d^3 p} = & \frac{d \, \tau_\text{fo}\, R_0^2 \,m_T \,p_T^2}{8\pi^2\,T_\text{fo}^2}\,K_1\left(\frac{m_T}{T}\right),\\
E \frac{dN_2}{d^3 p} = & \frac{d \, \tau_\text{fo}\,R_0^2\,m_T^2}{8 \pi^2\,T_\text{fo}}\,K_2\left(\frac{m_T}{T}\right),\\
E \frac{dN_3}{d^3 p} = & \frac{d \, \tau_\text{fo}\,R_0^2 \,m_T (m_T^2+T_\text{fo}^2)}{8\pi^2\,T_\text{fo}^4}\,K_1\left(\frac{m_T}{T}\right).
\end{split}
\end{equation}
In the simple model studied here, information about velocity correlations enters the spectrum only
via the coordinate-independent three numbers $G_u^{11}(\tau_\text{fo})$, $G_u^{yy}(\tau_\text{fo})$ 
and $G_T(\tau_\text{fo})$. For another model choice, the information may be slightly different. For 
instance, if one would replace the sharp freeze-out at $\tau_{\rm fo}$ by a decoupling at times $\tau$
around $\tau_{\rm fo}$, then the three numbers $G_u^{11}(\tau_\text{fo})$, $G_u^{yy}(\tau_\text{fo})$ 
and $G_T(\tau_\text{fo})$ would be replaced by averages over $\tau$. Irrespective of such 
model-dependent nuances, however, it is a generally known feature that the single particle spectrum (\ref{eq:CFfreezeout}) is only sensitive to space-time averages over the distribution $f(x,p)$ and therefore 
does not contain information about correlations between different space time points. 

\subsection{Generalization to identical two-particle correlations} 
As discussed in section~\ref{sec6}, the dependence of velocity correlations on the wave-numbers 
$k_1,k_2$ and $k_y$ allows for a detailed characterization of fluid dynamic behavior, including 
information about the dissipation of fluctuations and the manifestations of turbulence. The 
one-particle spectra discussed so far contain only the information (\ref{eq7.5}) about the 
correlations of fluid fields 
\begin{equation}
G_u^{ij}(\tau,x_1,x_2,y), \quad G_{uT}^j(\tau,x_1,x_2,y), \quad  G_T(\tau,x_1,x_2,y)
\label{eq:correlfunct123}
\end{equation}
at equal positions ($x_1=x_2=y=0$). Here, we point out that identical (Bose-Einstein) two-particle
correlation functions are linear functionals of \eqref{eq:correlfunct123} and may thus provide
information about the wave number dependence of velocity correlations.

Two-particle spectra for pairs of  identical bosons ($s_{B/F}=1$) or fermions ($s_{B/F}=-1$) 
of 4-momenta $p_A$, $p_B$ respectively, can be written as~\cite{Lisa:2005dd}
\begin{equation}
\begin{split}
E_A E_B & \frac{d N}{d^3 p_A d^3 p_B} = 
 \int (p_A)_\mu d\Sigma^\mu \; (p_B)_\nu d \Sigma^{\prime\nu} f(x,p_A) f(x^\prime,p_B) \\
& +s_{B/F} \int \tfrac{1}{2}(p_A+p_B)_\mu d\Sigma^\mu \;\tfrac{1}{2}(p_A+p_B)_\nu d\Sigma^{\prime\nu}
 e^{i(p_A-p_B)_\mu (x-x^\prime)^\mu} f(x,\tfrac{p_A+p_B}{2}) f(x^\prime,\tfrac{p_A+p_B}{2}).
\end{split}
\label{eq:twopartspect1}
\end{equation}
Here, the first term is what one would expect from kinetic theory for classical particles while the second term results from the quantum statistics of identical particles. Data about two-particle spectra are typically normalized by a 
mixed-event technique that corresponds to forming a normalized correlation function 
\begin{equation}
C(p_A,p_B) = \frac{E_A E_B \frac{dN}{d^3 p_A d^3 p_B}}{E_A \frac{dN}{d^3 p_A} E_B \frac{dN}{d^3 p_B}}.
\end{equation}

As it stands, equation \eqref{eq:twopartspect1} is valid for an event-specific realization of the velocity and
temperature fields. Experimental data are for event samples that correspond to averages over the
hydrodynamic fields. This amounts to replacing in equation \eqref{eq:twopartspect1} the products of 
phase-space distributions $f$ by the corresponding event averages 
$\langle f(x,p_A) f(x^\prime,p_B) \rangle$ and  $\langle f(x,\tfrac{p_A+p_B}{2})
 f(x^\prime,\tfrac{p_A+p_B}{2}) \rangle$. Paralleling the arguments employed for the calculation of 
 the one-particle spectra via \eqref{eq:Bdexpansioninfluct}, one should then expand the arguments
 of $\langle f(x,p_A) f(x^\prime,p_B) \rangle$  for small fluctuations around the event-averaged 
background fields. In general, the arguments of these averages depend on space-time
difference $x-x^\prime$, and they contain information about the relative position dependence
of the correlations~\eqref{eq:correlfunct123} in the fluid fields. 

In practice, the way in which information about ~\eqref{eq:correlfunct123} enters the two-particle 
correlation functions may depend significantly on model-specific choices for $f(x,p)$. For illustrative
purposes, we explore here the particularly simple Bjorken-like model without spatial constraints in the
transverse direction. We ignore all correlations involving the temperature field on the ground that
these are for a compressionless situation formally of higher order in the fluctuating velocities,
see Eq.~\eqref{eq:connectiondtov}. For the discussion in the following, we also neglect the rapidity-dependence 
of the correlation functions  thus eliminating many terms proportional to $G_u^{ij}$ with $i=y$ or $j=y$ (or both), 
as well as 
terms involving $G_{uT}^y$ (The assumption of a vanishing rapidity dependence is relaxed in Appendix \ref{sec:appA}). With these approximations,
we concentrate therefore on the correlation functions of velocities $G_u^{mn}$ with $m,n=1,2$. Due to rotational invariance in the transverse plane, the velocity correlation in Fourier space can be written in the form
\begin{equation}
\tilde G_u^{mn}(\tau, k_1,k_2,k_y) = 2 \pi \; \delta(k_y)
 \left[ \delta_{mn}\; g_1(\tau, k) + k_m k_n\; g_2(\tau,k) \right]\, ,
\label{eq:Gumn}
\end{equation}
with $k=\sqrt{k_1^2+k_2^2}$.
In terms of the pair momentum $P=\frac{1}{2}(p_A+p_B)$ and the relative momentum $q=p_A-p_B$,
the two-particle correlation function at mid-rapidity $\eta_A=\eta_B=0$
takes then the form (see appendix \ref{sec:appA} for details of the
derivation) 
\begin{equation}
\begin{split}
C(P,q) &  = 1 +  s_{B/F} \; (2\pi)^2 \delta^{(2)}(\vec q_T) \frac{1}{A_T} 
+ \frac{\vec P_T^2- \vec q_T^2/4}{A_T T_\text{fo}^2} g_1(\tau_\text{fo},0)
\\
& + s_{B/F}  \frac{(m_T^{AB})^2}{m_T^A m_T^B A_T T_\text{fo}^2} 
\left[ \vec P_T^2 \, g_1(\tau_\text{fo},q_T) + (\vec P_T\cdot \vec q_T)^2 g_2(\tau_\text{fo},q_T)\right]
 \frac{\left|K_1\left(\tfrac{m_T^{AB}}{T_\text{fo}}-i(m_T^A-m_T^B)\tau_\text{fo} \right) \right|^2}{K_1\left(\frac{m_T^A}{T_\text{fo}}\right) K_1\left(\frac{m_T^B}{T_\text{fo}}\right)},
\end{split}
\label{eq:Cwithfluct}
\end{equation}
with $\vec P_T=(P_1,P_2)$, $\vec q_T=(q_1,q_2)$ and $q_T=\sqrt{\vec q_T^2}$.

Realistic phase space distributions $f(x,p)$ have support in a finite transverse area $A_T$ only,
and this would lead to a fall-off of the correlation function to unity in the relative transverse momentum 
$\vec{q}_T$ on a scale of order $1/\sqrt{A_T}$. Often, this fall-off is parametrized by a Gaussian
ansatz in terms of  HBT radius parameters, so that the first two terms of \eqref{eq:Cwithfluct} would
take the form $C(P,q)  = 1 +  s_{B/F} \exp\left[- A_T\, q_T^2\right]$. For the simplified model of infinite 
transverse extension discussed here, this contribution is singular $\propto (2\pi)^2 \delta^{(2)}(\vec q_T)$ 
and we have written it in a formal way normalized to unity at $\vec q_T=0$. 

For the purpose of the following discussion, the transverse translational invariance of the present toy model
presents the technical advantage that the $\vec{q}_T$-dependence of the
correlation function \eqref{eq:Cwithfluct} is solely dependent on the wave number dependence of 
velocity correlations. Effects that could confound the interpretation of the  $\vec{q}_T$-dependence
in practice, such as effects from a finite transverse geometry and from transverse
(event-average) velocity gradients, are not included in the present model. Therefore, the following
 discussion allows us to illustrate how information
about velocity correlations enters two-particle correlation functions, but it limits our discussion of 
how such information could be disentangled from other effects in a 
phenomenologically relevant scenario.  Keeping this caveat in mind, we observe that
the first term in \eqref{eq:twopartspect1} can be viewed as an incoherent superposition of 
single-particle spectra and therefore does not furnish information that is not yet contained in
single-particle spectra. In contrast, the quantum-statistical second term $\propto s_{B/F}$
furnishes novel information about the wave number dependence of $g_1(\tau_\text{fo},q_T)$ 
and $g_2(\tau_\text{fo},q_T)$. 

We note as a curious aside that for a situation of fully developed two-dimensional turbulence, 
one can predict the form of $g_1(\tau,k)$ and $g_2(\tau,k)$ from the scaling theory of Kraichnan and Batchelor. In particular, comparing \eqref{eq:78}, \eqref{eq:tvsft} and \eqref{eq:Gumn} one can express $g_1$ and $g_2$
as a function of $E(t,k)$.
Moreover, from Batchelors scaling theory of freely decaying two-dimensional turbulence one finds then the following scaling in the inertial range
\begin{equation}
\begin{split}
g_1(\tau,k)  & = \frac{c}{\tau^{10/3} k^4},\\
g_2(\tau,k)  & = -\frac{c}{\tau^{10/3} k^6}\, ,
\end{split}
\end{equation}
with a common but unpredicted constant $c$ that reflects the absolute scale of velocity fluctuations. 
For the correlations function~\eqref{eq:Cwithfluct}, this results in an additive term with a very slow
power-law fall-off of the form $\propto \tfrac{1}{A_T\, T_\text{fo}^2\, q_T^4}$. Interestingly, if the velocity
correlations occurs on scales that are significantly smaller than the transverse extension $\sqrt{A_T}$ of 
the system, then the slow power-law $q_T$-dependence persists at 
relative momentum scales that are significantly larger than the typical scales $1/R$ set by HBT
radius parameters. It is an exciting possibility that a measurement
of a power-law $1/q_T^n$-dependence in two-particle correlation functions may provide 
a characteristic signature for turbulent distributions in the fluid dynamically evolved velocity fields 
of a heavy ion collision. We caution that 
the model discussed here is a simplified one; also, for intermediate Reynolds numbers one expects corrections
to the case of fully developed turbulence, see e.g. ~\cite{TurbulenceReviews}. 
What may persist in a phenomenologically realistic scenario, however, is the general idea that velocity 
correlations on small scales $l_T$ induce two-particle
correlations on large scales $q_T \sim 1/l_T$, and that these correlations are expected to be
governed by a power-law fall-off.


\section{Discussion and Conclusion}
Recent data analyses from RHIC and LHC have given support to arguments that soft hadron spectra 
may result from the fluid dynamic response to initial conditions with significant event-by-event fluctuations. 
Motivated by this suggestion, we have studied here how event-by-event fluctuations 
propagate on top of an event-averaged fluid dynamical background of Bjorken type.
The choice of a Bjorken background field is a simplification that retains essential elements of the 
expected fluid dynamical evolution of relativistic heavy ion collisions. As shown in the present paper,  
it allows for a particularly explicit, partly analytical discussion of the propagation of fluctuations in
an expanding fluid dynamic system. In particular, we have found for the case of a laminar evolution
explicit expressions for the attenuation or amplification of all fluid dynamic modes over the
time scale relevant in heavy ion collisions. And we have specified the general conditions
for non-linear effects in the dynamics of fluctuations, finding in particular that the late time
dynamics evolves towards an essentially two-dimensional system with turbulent behavior.
Here we discuss our main findings in more detail:

To discuss the propagation of fluctuations, one needs to specify first the nature of the
fluctuations that are propagated. Recent studies have focussed mainly on fluctuations
in the energy density (or, equivalently, entropy density), where the Glauber model provides
a phenomenologically supported basis for assuming local fluctuations on a particular transverse
scale. However, it had been pointed out already that fluctuations in the velocity field may be
present in the initial conditions for fluid dynamic evolution.  Velocity fluctuations could arise
from pre-equilibrium evolution, or they could be a natural consequence
of fluctuations in the primary interactions of elementary constituents. To the best of our knowledge,
there is no a priori argument that fluctuations in energy density dominate over
fluctuations in other fluid dynamic fields. Also, it requires studies allowing for all possible 
fluctuations to address the question whether and how fluctuations in velocity and 
energy density can be disentangled. 
In section~\ref{sec2}, we argued that a discussion of the fluid dynamic response to fluctuating initial conditions
should be based on a formulation that allows for fluctuations in {\it all} fluid dynamic fields.  
In particular, we supported with a model study the idea that fluctuations in the velocity fields
may carry significant vorticity. For a fluid with conserved charges such as baryon number and electric charge one should take fluctuations in these quantities into account, as well.

In general, a separation of fluid dynamical fields into background and fluctuations is not 
necessary. For instance, recent studies of event-by-event fluid dynamics propagate 
event samples of initial conditions numerically without separating fluctuations from background 
fields. In principle, such full fluid dynamical simulations allow to explore under the most versatile
model assumptions the fluid dynamic response to fluctuating initial conditions. But an explicit
mode-by-mode formulation of 
the dynamics of fluctuations around a fluid background field seems well-suited to study 
which fluctuating modes in the 
initial conditions can survive the strong dynamical evolution in a heavy ion collision unattenuated,
whether there are mechanisms that may amplify some modes, and which modes are 'filtered out'
by the medium due to dissipative effects in the fluid dynamic evolution. To address such questions
in a simplified framework that accounts for the main features of fluid expansion, we have formulated
in section~\ref{sec3} the fluid dynamical evolution  of local fluctuations around average fluid fields of 
Bjorken type. For this system, we have found a peculiar rescaling of the time variable,
$t \propto \tau^{4/3}$, that allowed us to write the relativistic fluid dynamic evolution of fluctuations in 
a form resembling a non-relativistic Navier-Stokes equation. 

If the Reynolds number of a fluid system is not too large, then important elements of the dynamics
may be understood in a linearized treatment. In section~\ref{sec4}, we observed 
that in this laminar case, fluctuations in vorticity decouple from sound modes and fluctuations
in energy density. For a parameter set that is characteristic for heavy ion collisions, we have then
studied how different modes are amplified or attenuated over the time scale of order 10 fm/c of a heavy ion collision. Remarkably, the longitudinal vorticity modes have an algebraic enhancement 
factor that can overcompensate  on this time scale the typical exponential decay due to dissipative
effects. Therefore, vorticity, if not present in the initial conditions, will not be generated as long as the dynamical evolution is laminar, and it is therefore unlikely to be generated in a sizable amount for 
small Reynolds numbers. However, if present in the initial conditions, some vorticity modes 
will be amplified significantly during the dynamical evolution. To the best of our knowledge,
this phenomenon has not been studied yet in numerical simulations, and it would be interesting
to see how it manifests itself in the presence of other background fields.  In addition to the vorticity
modes, we have also explored the propagation of sound modes that result from initial fluctuations. 
In general, we find that fluctuations of sufficiently long wave-length pass unattenuated over time 
scales relevant for heavy ion collisions, while short wave-lengths that reflect finer structures in 
the fluctuating initial conditions, are dissipated on shorter length scales.  There is also a characteristic
difference between transverse components, and the components that propagate in the 
longitudinal direction in which the system expands according to Bjorken's model. 

Outside the regime of validity of a linearized treatment, the discussion of solutions of fluid 
dynamics is very complicated and typically requires numerical techniques. For our problem of
fluctuations around Bjorken flow, we are in the special and fortunate case that we can relate the
full relativistic dynamics of fluctuations in rescaled coordinates to a non-relativistic Navier-Stokes
equation. This allows us to discuss the possibility of a turbulent evolution in terms of those
concepts and parametric estimates that have proven useful in characterizing turbulent phenomena
of non-relativistic systems. In section~\ref{sec5}, we introduce both a longitudinal and a transverse
Reynolds number to characterize the non-linear dynamics of fluctuations on top of a Bjorken background
field that shows strong dynamical expansion only in the longitudinal direction. We find in particular that
the late time dynamics will evolve a large set of initial conditions into a regime where the dynamical
evolution is effectively two-dimensional, and where the transverse Reynolds number can be sizable.
This indicates a window for a two-dimensional, non-linear evolution towards turbulent behavior. 
Motivated by this observation, we have summarized in section~\ref{sec6} characteristic features
of turbulent behavior, detailing the differences between the cases of two-dimensional and
three-dimensional evolution. We recall from this discussion in particular that 
in three dimensions, fluid kinetic energy thermalizes typically by dissipating into vorticity modes
of increasing wave number, i.e. decreasing length scales. As first observed by Kraichnan, this 
mechanism is not possible for a two-dimensional fluid system where one
finds an inverse cascade: kinetic energy gets propagated to larger length scales. 

The transverse Reynolds numbers estimated in section~\ref{sec5} do not support the assumption that 
heavy ion collisions create systems with fully developed turbulence. However, our discussion in
section~\ref{sec5} shows that realistic Reynolds numbers are not small enough to neglect 
non-linear effects in the dynamical evolution. 
Therefore, while we have no rationale to expect that the correlation functions of fluid dynamic fields
generated in heavy ion collisions satisfy the scaling laws of Kolmogorov's theory of fully developed
turbulence, we do expect that a non-linear dynamics that can be regarded as the onset of 
turbulent behavior may result in interesting (power-law) wave-number dependences of correlations
of fluid fields. By supplementing a standard blast-wave model with event-by-event fluctuations in
fluid fields, we have established in Sect.~\ref{sec7} how such fluid dynamic correlation functions manifest
themselves in one- and identical two-particle spectra. For one-particle spectra, we find that fluctuations
are a confounding factor in interpreting the $m_T$-dependence of spectra in a fluid dynamic scenario. 
For two-particle correlations, we observe a dependence that may provide an experimentally 
accessible signature for the onset of turbulence in heavy ion collisions. The observation is that 
identical two-particle correlations at large relative momentum are sensitive to spatial
scales that are much smaller than the transverse size of the particle producing source. If two-point
velocity correlations show turbulent behavior on these small scales then this will translate into a
characteristic power-law tail of identical two-particle correlations at large relative momentum.  

Let us close with a short outlook of how observations made in this study could be
pursued further. Our study was largely motivated by the question of how 
initial conditions with significant velocity fluctuations (in the solenoidal an in the irrotational part) 
propagate fluid dynamically.  Here, full fluid dynamic simulations including initial velocity 
fluctuations could provide further insight, for instance by evolving fluctuations
around average fluid fields with more realistic transverse dependencies, and by quantifying to what 
extent initial velocity fluctuations could contribute to the observed azimuthal asymmetries in 
momentum space. Full fluid dynamic simulations including initial velocity fluctuations could also 
allow for a detailed characterization of how the scale dependence of fluid correlation functions of 
the type \eqref{eq:correlfunct123} builds up during the fluid dynamic evolution. This would provide 
in particular insight into the question on which time scales
and over which range of wave vectors correlation functions of fluid fields may develop power-law
dependences that can be regarded as precursors of turbulent phenomena~\footnote{We note in this context
that the time scale on which non-linear contributions start to matter in the evolution of fluctuations may
depend sensitively on the size and scale of the fluctuating initial condition.}. Also, the semi-analytical 
approach used in the present work may be pursued further. In the present work, we have seen how
single vorticity modes, and sound modes are amplified or filtered out by the dynamical evolution, 
depending on their wave-number. We plan to investigate, whether a similar understanding can be gained for 
a more realistic background field by expanding fluctuating fluid fields in terms of appropriate sets of functions so that
one can calculate explicitly how the fluid dynamic
evolution mixes different components in the evolution. We expect that such studies could provide
an intuitive understanding e.g. of the time scales on which density fluctuations feed sound waves,
or on which vorticity modes cascade to other scales. This may help significantly in the interpretation
of the complex fluid dynamic phenomena that we expect to find realized in heavy ion collisions.

\begin{appendix}
\section{Explicit expressions for the two-particle spectrum}
\label{sec:appA}

Here, we provide further details about the calculation of the two-particle correlation function~\eqref{eq:Cwithfluct},
and how the calculation of this correlation function could be generalized to include the rapidity dependence
of  velocity correlations, as well as effects of temperature fluctuations. In general,  we represent fluctuations in the
fluid dynamic fields in Fourier space according to
\begin{equation}
\begin{split}
&T(\tau,x_1,x_2,y)-T_\text{fo} = 
\int \frac{d^3 k}{(2\pi)^3} \, e^{i(k_1x_1+k_2x_2+k_y y)}\; \tilde T(\tau,k_1,k_2,k_y),\\
&u^j(\tau,x_1,x_2,y) =
\int \frac{d^3 k}{(2\pi)^3} \, e^{i(k_1x_1+k_2x_2+k_y y)}\; \tilde u^j(\tau,k_1,k_2,k_y).
\end{split}
\end{equation}
For the first term of the two-particle spectrum \eqref{eq:twopartspect1}, we find then the following contribution ($m_T^A$, $m_T^B$, $\eta_A$, and $\eta_B$ are the transverse masses and rapidities of particles $A$ and $B$)
%
\begin{equation}
\begin{split}
&\frac{d^2 \tau_\text{fo}^2 m_T^A m_T^B }{(2\pi)^6} \int \frac{d^3k}{(2\pi)^3} \frac{d^3k^\prime}{(2\pi)^3}\\
&\times \int dx_1 dx_2 dy \,\text{cosh}(\eta_A-y) e^{i(k_1x_1+k_2x_2+k_y y)} e^{-\frac{m_T^A}{T_\text{fo}}\text{cosh}(\eta_A-y)}\\
&\times {\bigg \langle} \left[ \frac{m_T^A}{T_\text{fo}} \tilde T(k) \text{cosh}(\eta_A-y) + \frac{(p_A)_1 \tilde u^1(k) + (p_A)_2 \tilde u^2(k)}{T_\text{fo}} + \frac{m_T^A\, \tau\, \tilde u^y(k)}{T_\text{fo}} \text{sinh}(\eta_A-y) \right] \\
&\times \int dx_1^\prime dx_2^\prime dy^\prime \text{cosh}(\eta_B-y^\prime) e^{i(k_1^\prime x_1^\prime+k_2^\prime x_2^\prime+k_y^\prime y^\prime)} e^{-\frac{m_T^B}{T_\text{fo}}\text{cosh}(\eta_B-y^\prime)}\\
&\times \left[ \frac{m_T^B}{T_\text{fo}} \tilde T(k^\prime) \text{cosh}(\eta_B-y^\prime) + \frac{(p_B)_1 \tilde u^1(k^\prime) + (p_B)_2 \tilde u^2(k^\prime)}{T_\text{fo}} + \frac{m_T^B\, \tau\, \tilde u^y(k^\prime)}{T_\text{fo}} \text{sinh}(\eta_B-y^\prime) \right] {\bigg \rangle}.
\end{split}
\label{eq:fluctxxprime1}
\end{equation}
%
For an infinite extension in the transverse plane it is easy to perform the integrals over $x_1,x_2, x_1^\prime, x_2^\prime$. The resulting Dirac distributions can be used to perform the integrals over $k_1,k_2,k_1^\prime, k_2^\prime$. Also, the integrals over $y$ and $y^\prime$ can be done analytically.
Transforming then back to Fourier space, one finds for the first term of \eqref{eq:twopartspect1} the expression
\begin{equation}
\begin{split}
\frac{d^2 \tau_\text{fo}^2 m_T^A m_T^B A_T }{(2\pi)^6} &\int \frac{d k_y}{2\pi} e^{i k_y (\eta_A-\eta_B)}
 {\bigg [} \frac{m_T^A m_T^B}{T_\text{fo}^4} \tilde G_T(\tau,0,0,k_y) E_1\left(\tfrac{m_T^A}{T_\text{fo}},k_y\right) E_1\left(\tfrac{m_T^B}{T_\text{fo}},-k_y\right)\\
&+ \frac{ (\vec p_A)_T (\vec p_B)_T}{T_\text{fo}^2} G_u^{11}(\tau,0,0,k_y) E_0\left(\tfrac{m_T^A}{T_\text{fo}},k_y\right) E_0\left(\tfrac{m_T^B}{T_\text{fo}},-k_y\right)\\
& + \frac{\tau_\text{fo}^2 m_T^A m_T^B }{T_\text{fo}^2} G_u^{yy}(\tau,0,0,k_y) E_2\left(\tfrac{m_T^A}{T_\text{fo}},k_y\right) E_2\left(\tfrac{m_T^B}{T_\text{fo}},-k_y\right) {\bigg ]}\, ,
\end{split}
\end{equation}
where the functions $E_i$ denote linear combinations of Bessel functions of the second kind, 
\begin{equation}
\begin{split}
E_0(x,q) & = K_{1+iq}(x) + K_{1-iq}(x),\\
E_1(x,q) & = \frac{1}{2} K_{2+iq}(x) +\frac{1}{2} K_{2-iq}(x) + K_{iq}(x),\\
E_2(x,q) & = \frac{1}{2} K_{2+iq}(x) - \frac{1}{2} K_{2-iq}(x)\, .
\end{split}
\end{equation}
For $k_y = 0$, this expression reduces to the first two terms in the first line of \eqref{eq:Cwithfluct}. 

Let us now consider the second term $\sim s_{B/F}$ in Eq.\ \eqref{eq:twopartspect1}. This term is of the 
form of a single integral over the freeze-out volume
\begin{equation}
\int \frac{\tfrac{1}{2}(p_A+p_B)_\mu d\Sigma^\mu}{(2\pi)^3} e^{i(p_A-p_B)_\mu x^\mu} f(x,\tfrac{p_A+p_B}{2})
\label{eq:HBTterm}
\end{equation}
times its complex conjugate. Expanding it up to terms that are quadratic in the fluctuations, one obtains two
distinct contributions, In one case,  the bilinear terms in the fluctuating fields are written 
at different points $x$, $x^\prime$, in the other case they are both taken at the same point. 
We consider both cases separately.
We work in the following with average pair momentum $P^\mu=\frac{1}{2}(p_A^\mu+p_B^\mu) = 
(m_T^{AB} \text{cosh}\, \eta_{AB}, P^1, P^2, m_T^{AB}\text{sinh}\, \eta_{AB})$ and we define
a transverse mass defined $m_T^{AB}=\sqrt{(P^0)^2-(P^3)^2}$ and a 
rapidity $\eta_{AB} = \text{arctanh}(P^3/P^0)$. This allows us to write 
$q_\mu x^\mu = -[m_T^A \text{cosh}(\eta_A-y)-m_T^B \text{cosh}(\eta_B-y)]\tau + \vec q_T \vec x_T$ with $\vec q_T=(q_1,q_2)$ and $\vec x_T=(x_1,x_2)$.

The contribution to the second term in Eq.\ \eqref{eq:twopartspect1} that is quadratic in fluctuations at
the same space-time point, can then be written as
\begin{equation}
\begin{split}
& \frac{d\, \tau_\text{fo}\, m_T^{AB}}{(2\pi)^3} \int dx_1 dx_2 dy \; \text{cosh}(\eta_{AB}-y) 
 e^{-i\tau \left[ m_T^A \text{cosh}(\eta_A-y)-m_T^B \text{cosh}(\eta_B-y) \right]\tau_\text{fo}}
 e^{i \vec q_T \vec x_T}\; e^{-\frac{m_T^{AB}}{T_\text{fo}}\text{cosh}(\eta_{AB}-y)}\\
&\times {\bigg [} 1+ \frac{1}{2\,T_\text{fo}^2}\vec P_T^2 \,G_u^{11}(\tau) 
+\frac{\tau^2 (m_T^{AB})^2}{2\,T_\text{fo}^2} \text{sinh}^2(\eta_{AB}-y) \,G_u^{yy}(\tau) \\
&\qquad +\left( \frac{(m_T^{AB})^2 \text{cosh}^2(\eta_{AB}-y)}{2\,T_\text{fo}^2}-\frac{m_T^{AB} \text{cosh}(\eta_{AB}-y)}{2\,T_\text{fo}^3} \right)  G_T(\tau)  {\bigg ]}.
\end{split}
\label{eq:twopartspect2}
\end{equation}
Here, the integral over the transverse coordinates leads to a factor $(2\pi)^2 \delta^{(2)}(\vec q_T)$ (where $(2\pi)^2 \delta^{(2)}(0) = A_T$ is understood). This is a consequence of the fact that in the present model the transverse
extension of the particle emitting source is not limited. At mid-rapidity, $\eta_A=\eta_B=\eta_{AB}=0$, the term \eqref{eq:twopartspect2} simplifies to
\begin{equation}
\begin{split}
& \frac{d\, \tau_\text{fo}\, m_T^{AB}}{(2\pi)^3} (2\pi)^2 \delta^{(2)}(\vec q_T)
{\bigg [} \left(1+\frac{p_T^2}{2T_\text{fo}^2}G_u^{11}(\tau)+\frac{m_T^2+T_\text{fo}^2}{2T_\text{fo}^2} G_T(\tau)\right) 2\, K_1\left(\tfrac{m_T}{T_\text{fo}}\right)
+ \frac{\tau^2 m_T}{2T_\text{fo}} G_u^{yy}(\tau)\, 2\, K_2\left(\tfrac{m_T}{T_\text{fo}}\right) {\bigg ]}.
\end{split}
\end{equation}
With the approximations used in section~\ref{sec7}, this expression 
reduces to the last terms in the first line of \eqref{eq:Cwithfluct}.

We now turn to the contribution $\sim s_{B/F}$ where the fluctuating fields have different space-time argument. This term is of similar structure as \eqref{eq:fluctxxprime1} and reads
\begin{equation}
\begin{split}
&\frac{d^2 \tau_\text{fo}^2 (m_T^{AB})^2}{(2\pi)^6} \int \frac{d^3k}{(2\pi)^3} \frac{d^3k^\prime}{(2\pi)^3}\\
&\times \int dx_1 dx_2 dy \,\text{cosh}(\eta_{AB}-y) e^{i(k_1x_1+k_2x_2+k_y y)} e^{-\frac{m_T^{AB}}{T_\text{fo}}\text{cosh}(\eta_{AB}-y)} \\
&\times e^{-i[m_T^A \text{cosh}(\eta_A-y)-m_T^B \text{cosh}(\eta_B-y)]\tau_\text{fo}} \; e^{i\vec q_T \vec x_T}\\
&\times {\bigg \langle} \left[ \frac{m_T^{AB}}{T_\text{fo}} \tilde T(k) \text{cosh}(\eta_{AB}-y) + \frac{P_1 \tilde u^1(k) + P_2 \tilde u^2(k)}{T_\text{fo}} + \frac{\tau_\text{fo}\,m_T^{AB}\,  \tilde u^y(k)}{T_\text{fo}} \text{sinh}(\eta_{AB}-y) \right] \\
&\times \int dx_1^\prime dx_2^\prime dy^\prime \text{cosh}(\eta_{AB}-y^\prime) e^{i(k_1^\prime x_1^\prime+k_2^\prime x_2^\prime+k_y^\prime y^\prime)} e^{-\frac{m_T^{AB}}{T_\text{fo}}\text{cosh}(\eta_{AB}-y^\prime)}\\
&\times e^{i[m_T^A \text{cosh}(\eta_A-y^\prime)-m_T^B \text{cosh}(\eta_B-y^\prime)]\tau_\text{fo}} \; e^{-i\vec q_T \vec x^\prime}\\
&\times \left[ \frac{m_T^{AB}}{T_\text{fo}} \tilde T(k^\prime) \text{cosh}(\eta_{AB}-y^\prime) + \frac{P_1 \tilde u^1(k^\prime) + P_2 \tilde u^2(k^\prime)}{T_\text{fo}} + \frac{\tau_\text{fo}\,m_T^{AB}\,  \tilde u^y(k^\prime)}{T_\text{fo}} \text{sinh}(\eta_{AB}-y^\prime) \right] {\bigg \rangle}.
\end{split}
\label{eq:fluctxxprime2}
\end{equation}
The integrals over the transverse coordinates $x_1,x_2,x_1^\prime, x_2^\prime$ leads to factors $(2\pi)^2 \delta^{(2)}(\vec q_T+\vec k_T)$ and $(2\pi)^2 \delta^{(2)}(\vec q_T- \vec k_T^\prime)$, respectively. These can be used to perform the integrals over the transverse components of $k$ and $k^\prime$. For simplicity we concentrate again on particles at mid-rapidity, $\eta_A=\eta_B=\eta_{AB}=0$. It is then straight-forward to perform the integrals over $y$ and $y^\prime$ 
In terms of the fluid dynamic correlation functions, the result can then be written as 
\begin{equation}
\begin{split}
& \frac{d^2 \tau_\text{fo}^2 (m_T^{AB})^2 A_T}{(2\pi)^6} \int \frac{d k_y}{2\pi}
 {\Bigg [} \frac{(m_T^{AB})^2}{T_\text{fo}^2} G_T(\tau,\vec q_T,k_y) \left| E_1\left( \tfrac{m_T^{AB}}{T_\text{fo}}-i(m_T^A-m_T^B)\tau_\text{fo}, k_y \right) \right|^2 \\
&+ \sum_{m,n=1}^2 \frac{P_m P_n}{T_\text{fo}^2} G_u^{mn}(\tau,\vec q_T,k_y) \left| E_0\left( \tfrac{m_T^{AB}}{T_\text{fo}}-i(m_T^A-m_T^B)\tau_\text{fo}, k_y \right) \right|^2\\
&+\frac{\tau_\text{fo}^2 (m_T^{AB})^2}{T_\text{fo}^2} G_u^{yy}(\tau,\vec q_T,k_y) \left| E_2\left( \tfrac{m_T^{AB}}{T_\text{fo}}-i(m_T^A-m_T^B)\tau_\text{fo}, k_y \right) \right|^2\\
&+\sum_{n=1}^2 \frac{P_n}{T_\text{fo}^2} {\bigg \{} G_{uT}^n(\tau,\vec q_T,k_y)  E_1\left( \tfrac{m_T^{AB}}{T_\text{fo}}-i(m_T^A-m_T^B)\tau_\text{fo}, k_y \right)  E_0\left( \tfrac{m_T^{AB}}{T_\text{fo}}+i(m_T^A-m_T^B)\tau_\text{fo}, -k_y \right) + c.c. {\bigg \}} \\
&+\frac{\tau_\text{fo}(m_T^{AB})^2}{T_\text{fo}^2} {\bigg \{} G_{uT}^y(\tau,\vec q_T,k_y)  E_1\left( \tfrac{m_T^{AB}}{T_\text{fo}}-i(m_T^A-m_T^B)\tau_\text{fo}, k_y \right) E_2\left( \tfrac{m_T^{AB}}{T_\text{fo}}+i(m_T^A-m_T^B)\tau_\text{fo}, -k_y \right) + c.c. {\bigg \}} \\
&+\sum_{n=1}^2 \frac{\tau_\text{fo}\, m_T^{AB}\, P_n}{T_\text{fo}^2} {\bigg \{} G_{u}^{ny}(\tau,\vec q_T,k_y)  E_0\left( \tfrac{m_T^{AB}}{T_\text{fo}}-i(m_T^A-m_T^B)\tau_\text{fo}, k_y \right) \\
& \qquad \times E_2\left( \tfrac{m_T^{AB}}{T_\text{fo}}+i(m_T^A-m_T^B)\tau_\text{fo}, -k_y \right) + c.c. {\bigg \}}
{\Bigg ]}.
\end{split}
\label{eq:x12}
\end{equation}
We note that \eqref{eq:x12} contains information about the full fluid dynamic correlation functions in 
momentum space. Under the assumptions made in section~\ref{sec7}, this expression reduces to
the last line of \eqref{eq:Cwithfluct}.
\end{appendix}


\medskip
\section*{Acknowledgments}
\medskip
We thank R.~Baier and J.~Casalderrey-Solana for useful discussions. S.~F.~acknowledges financial support by DFG under contract FL 736/1-1.


\end{document}